\documentclass[preprint]{JHEP3} 

\usepackage{epsfig}
\usepackage{amsfonts}
\usepackage{amssymb,amsmath}

\newcommand{\mbf}[1]{{\boldsymbol {#1} }}
\newcommand{\complex}{{\mathbb C}} 
\newcommand{\zed}{{\mathbb Z}} 
\newcommand{\nat}{{\mathbb N}} 
\newcommand{\real}{{\mathbb R}} 
\newcommand{\rat}{{\mathbb Q}} 
\newcommand{\torus}{{\mathbb T}}
\def\e{{\,\rm e}\,}

\def\ii{{\,{\rm i}\,}}
\def\dd{{\rm d}}
\def\Li{{\rm Li}}

\def\beq{\begin{equation}}
\def\bee{\begin{equation}}
\def\eeq{\end{equation}}
\def\bea{\begin{eqnarray}}
\def\eea{\end{eqnarray}}
\def\bd{\begin{displaymath}}
\def\ed{\end{displaymath}}

\newcommand{\Cint}{\int\kern-10.5pt-\kern7pt}

\newcommand{\PP}{{\mathbb{P}}}

\newcommand{\be}{\begin{equation}}
\newcommand{\ee}{\end{equation}}

\newcommand\fverb{\setbox\pippobox=\hbox\bgroup\verb}
\newcommand\fverbdo{\egroup\medskip\noindent%
                        \fbox{\unhbox\pippobox}\ }
\newcommand\fverbit{\egroup\item[\fbox{\unhbox\pippobox}]}
\newbox\pippobox

\allowdisplaybreaks

\title{Topological strings and large $\mbf N$ phase transitions II:\\
  Chiral expansion of $\mbf q$-deformed Yang-Mills theory}
\author{Nicola Caporaso$^{(a)}$, Michele Cirafici$^{(b)}$, Luca
Griguolo$^{(c)}$,
 Sara Pasquetti$^{(c)}$,  Domenico~Seminara$^{(a)}$ and
 Richard~J.~Szabo$^{(b)}$ \\
$^{(a)}$ Dipartimento di Fisica, Polo Scientifico Universit\`a di
Firenze,\\ INFN Sezione di Firenze
Via  G. Sansone 1, 50019 Sesto Fiorentino, Italy\\
$^{(b)}$ Department of Mathematics and
Maxwell Institute for Mathematical Sciences,\\
Heriot-Watt University, Colin Maclaurin Building, Riccarton, Edinburgh
EH14 4AS, UK\\
$^{(c)}$ Dipartimento di  Fisica, Universit\`a  di Parma,
INFN Gruppo Collegato di Parma,\\
Parco Area delle Scienze 7/A, 43100 Parma, Italy\\
\email{caporaso@fi.infn.it , M.Cirafici@ma.hw.ac.uk ,
griguolo@fis.unipr.it , pasquetti@fis.unipr.it ,
seminara@fi.infn.it , R.J.Szabo@ma.hw.ac.uk } }

\received{\today}               

\accepted{\today}               
\preprint{ {\tt HWM-05-22} \ \ {\tt EMPG-05-16}
\\ \hepth{0511043}} 

\date{data}
\abstract{We continue our study of the large $N$ phase transition in
  $q$-deformed Yang-Mills theory on the sphere and its role in
  connecting topological strings to black hole entropy. We study
  in detail the chiral theory defined in terms of uncoupled single $U(N)$
representations at large $N$ and write down the resulting
partition function by means of the topological vertex. The
emergent toric geometry has three K\"ahler parameters,
one of which corresponds to the expected fibration over
$\PP^1$. By taking a suitable double-scaling limit
we recover the chiral Gross-Taylor string expansion. To analyse the
phase transition we construct a matrix model which describes the
chiral gauge theory. It has three
distinct phases, one of which should be described by the
closed topological string expansion. We
verify this expectation by explicit comparison between the
matrix model and the chiral topological string free energies. We also show that
the critical point in the pertinent phase of the matrix model corresponds to
a divergence of the topological string perturbation series.}


\begin{document}

\section{Introduction}

The study of black holes in string theory has received a renewed
attention after the Ooguri-Strominger-Vafa
conjecture~\cite{Ooguri:2004zv} which proposes a connection between the
entropy of four dimensional BPS black holes and topological string
amplitudes. A possible way to construct such a black hole is
through a Calabi-Yau compactification
 of Type~IIA superstring theory which yields $\mathcal{N}=2$
 supergravity in four dimensions. Four dimensional black hole
 microstates are interpreted as bound states of D-branes wrapping
 cycles of the internal manifold
and the microscopic entropy agrees at leading order with the
Bekenstein-Hawking entropy computed 
 in the corresponding effective field theory
 \cite{Strominger:1996sh}. Rather remarkably, 
 this relation holds beyond leading order when we include corrections
 to the Bekenstein-Hawking area 
 law that arise from higher derivative F-term interactions in the
 effective theory
 \cite{LopesCardoso:1998wt}--\cite{LopesCardoso:1999ur}
 (see~\cite{Mohaupt:2000mj} for a review).

On the other hand, the physics of the F-term in the four dimensional
effective field theory is captured by a twisted sigma-model defined on
the Calabi-Yau manifold. The genus
 zero free energy computes the effective action for the vector
 multiplets (in the Type~IIA setting)
up to two derivatives, while higher genus topological string
amplitudes encode information about higher derivative couplings
between the curvature and the graviphoton field
 strength  \cite{Antoniadis:1993ze,Bershadsky:1993cx}
 (see~\cite{Neitzke:2004ni}--\cite{Marino:2004eq} for reviews).

From this relation it have been argued~\cite{Ooguri:2004zv} that the
partition function of the black hole at the attractor point
\cite{Ferrara:1995ih,Strominger:1996kf} is equal to the modulus
squared of the topological string vacuum amplitude, $Z_{\rm BH} =
|Z_{\mathrm{top}}|^2$. This conjecture has been addressed and refined in
\cite{Dabholkar:2004yr}--\cite{Dabholkar:2005dt}.
A concrete realization of this proposal has been advanced in
\cite{Vafa:2004qa} and subsequently in \cite{Aganagic:2004js} for
local threefolds $X_p$ which are fibred over a compact curve $\Sigma_g$ of
genus $g$. In this case a black hole can be engineered as bound states
of $N$ D4-branes wrapping a non-compact four-cycle, and an arbitrary
 number of D-branes wrapping $\Sigma_g$ along with D0-branes. This
 configuration realizes a four dimensional BPS black hole whose
 microstates are described by a mixed ensemble of fixed magnetic
 D4-brane charge $N$ and electric chemical potentials.
 Bound states of D-branes are counted by the corresponding gauge field
 configurations excited on the worldvolume of the D4-brane. These in
 turn localize to a deformed version of $U(N)$ Yang-Mills theory on
 the Riemann surface $\Sigma_g$, the $q$-deformed
 Yang-Mills theory introduced
 in~\cite{Klimcik:1999kg,Aganagic:2004js}. In the large $N$ limit,
 this theory factorizes into a chiral and antichiral part, like its
 undeformed cousin QCD$_2$
 \cite{Gross:1992tu}--\cite{Gross:1993yt},
corresponding to the holomorphic and antiholomorphic structure
of $|Z_{\mathrm{top}}|^2$.
 The topological string amplitude $Z_{\mathrm{top}}$ on these geometries
has been computed recently in \cite{Bryan:2004iq}.

In the following we will focus on the case where the local threefold
$X_p$ is fibered over a two-sphere. This paper is a companion of
\cite{Caporaso:2005ta} where the gauge theoretical aspects of the
correspondence were analysed. Here
 we will study in detail the emergence of the topological string theory
 from the strong coupling phase of the chiral two-dimensional
 $q$-deformed gauge theory at large $N$.
A crucial outcome of our investigation is a path towards a topological
sigma-model description underlying the Gross-Taylor string expansion
\cite{Gross:1992tu}--\cite{Gross:1993yt}
of ordinary Yang-Mills theory. While at zero coupling (where QCD$_2$
is a topological field theory) such a reformulation
exists~\cite{Cordes:1994sd}, at finite area the full description in
terms of topological string theory is
problematic. We present a potential way to overcome these difficulties
which relies on viewing QCD$_2$ as a particular limit of the
$q$-deformed gauge theory. As we will show, the latter model admits a
well-defined interpretation as a topological string theory from which
one can in principle extract the precise string theory underlying
QCD$_2$. Moreover, as the topological strings live in higher target
space dimensions this approach has the potential of extending the
two-dimensional Gross-Taylor string description to QCD$_4$.

Many of the features of two-dimensional Yang-Mills theory on a sphere,
such as its non-trivial instanton driven phase transition in the large
$N$ limit \cite{Douglas:1993ii,Gross:1994mr}, are preserved by the
$q$-deformation and were thoroughly addressed
in~\cite{Caporaso:2005ta,Arsiwalla:2005jb,Jafferis:2005jd}. In this
paper we will focus on the chiral-antichiral decomposition of
$q$-deformed Yang-Mills theory,
which according to \cite{Vafa:2004qa,Aganagic:2004js} is the
chiral (antichiral) sector that should be related to the holomorphic
(antiholomorphic) topological string amplitudes. Similarly
 to what has been observed in \cite{Taylor:1994zm,Crescimanno:1994eg} for the
 undeformed theory, we will show that the chiral sector has a non-trivial
 phase structure that plays a key role in the comparison to string
 theory.

On the other hand, the threefold $X_p$ is one of the simplest and best
studied examples of a toric variety. The topological vertex
\cite{Aganagic:2003db} is a useful tool for computing topological string
 amplitudes in these geometries. A toric threefold is characterized by a graph
 where trivalent vertices are glued together and the edges signal the
 degeneration of a $\mathbb{T}^2$ fibration. Topological string
 amplitudes can be reduced (roughly by an appropriate placing of
 brane-antibrane pairs) to those on
patches of trivial topology. The topological vertex is the building block
of this construction. We will use this formalism to make explicit contact
 with the chiral $q$-deformed gauge theory at large $N$. In this setup
 we can make a remarkable check of the proposal by exhibiting a
relation between the Hurwitz numbers that compute the combinatorics
 of branched covering maps to $\mathbb{P}^1$ and the relevant Gromov-Witten
 invariants that ``count'' in the appropriate way
 the holomorphic curves embedded in the Calabi-Yau manifold. Moreover, the
 string theory exhibits its origin as a gauge theory
 through the finiteness of the radius of convergence of its
 perturbative expansion.

This paper is organized as follows. Section 2 is devoted to exploring the
 relationship between the large $N$ chiral $q$-deformed gauge theory
 partition function and the
 closed topological string amplitude on the local threefold $X_p$. We
 rewrite the free energy of the chiral theory in the large $N$ limit
 in terms of (generalized) Gromov-Witten invariants and show
 that its undeformed limit agrees perfectly with
 the conventional chiral QCD$_2$ result. We rewrite the chiral
 partition function in terms of the topological vertex as a
 topological string amplitude and analyse the Gromov-Witten invariants of
 the corresponding toric scheme $\hat{X}_p$. As expected, the related
 Gopakumar-Vafa invariants of $\hat{X}_p$ are integral for all $p$. We
 also use the undeformed limit to derive an asymptotic localization
 formula for the Gromov-Witten invariants in terms of Hodge integrals.

Section 3 is devoted to the analysis of the large $N$ phase structure
of the chiral gauge theory by means of a matrix model, and the recovery
of the topological string theory in the strong coupling phase. Our
saddle-point equations consistently reduce to the
Crescimanno-Taylor equations \cite{Crescimanno:1994eg} in the
undeformed limit and we find a qualitatively similar phase diagram to that of
chiral QCD$_2$. In particular, two phase transitions occur and we show
that the one-cut solution in the strong coupling phase agrees with the
expectations from topological string theory.

Finally, in Section 4 we examine the analytic properties of the perturbative
 expansion of the string partition function. We find a
finite radius of convergence that corresponds to the critical points
 of the phase transition. We compare the value of the critical point
 obtained in this way with both the
 numerical results of the matrix model analysis and with the exact result
 for the full coupled gauge theory found in \cite{Caporaso:2005ta}. We
 also speculate that the finite radius of convergence of the
 topological string amplitude may have an interpretation as a sort of
 Hagedorn transition. Section~5 contains some concluding remarks and
 we collect various technical details in four appendices at the end of
 the paper.

\section{$\mbf q$-deformed Yang-Mills theory on $\mbf{S^2}$ and
  closed topological strings\label{qdefS2}}

The conjectured relation in~\cite{Ooguri:2004zv} links
four-dimensional black holes with topological strings. Consider
Type~IIA string theory on the local Calabi-Yau threefold
\beq
X_p={\cal O}(-p)\oplus{\cal O}(p-2)~\longrightarrow~\mathbb{P}^1 \ ,
\label{Xfib}\eeq
where $p\in\zed$ and ${\cal O}(m)$ is the canonical holomorphic line bundle
over $\PP^1$ of degree $m$. The conjecture predicts that the black
hole partition function $Z_{\rm BH}$ is related for
large charges to the perturbative vacuum amplitude
$Z_{\rm top}$ for topological strings on $X_p$ by $Z_{\rm BH}=|Z_{\rm
  top}|^2$. Because an exact microscopic
computation~\cite{Aganagic:2004js} shows that $Z_{\rm BH}$
coincides with the partition function ${\cal Z}_{\rm YM}^q$ of
$q$-deformed $U(N)$ Yang-Mills theory on the base $S^2\cong\PP^1$
of the fibration (\ref{Xfib}), a natural check of the conjecture
is to verify that ${\cal Z}_{\rm YM}^q=Z_{\rm
top}\,\overline{Z}_{\rm top}$ in the large $N$ limit. In this
section we shall analyse this last relation by considering in
detail the large $N$ {\it chiral} expansion of the $q$-deformed
gauge theory. We will find the deformed analog of chiral QCD$_2$
and exhibit its precise relation with the closed topological
string amplitude on the Calabi-Yau threefold $X_p$. Unless
explicitly stated otherwise, we will assume that $p>2$ as this is
when the large $N$ phase transition occurs.

\subsection{Large $N$ expansion\label{LargeNexp}}

The partition function of $q$-deformed Yang-Mills theory on the sphere
$S^2$ is given by
\beq {\cal Z}^q_{\rm
YM}=\sum_R\,{\rm dim}_q(R)^2~q^{\frac{p}{2}\,C_2(R)} \ ,
\label{Parti}\eeq
where the sum runs over all irreducible representations of the $U(N)$
gauge group, $q=\e^{-g_s}$, $C_2(R)$ is the quadratic Casimir
invariant of $R$, and the quantum dimension is defined as
\begin{equation}
{\rm dim}_q(R)=\prod_{1\leq i<j\leq
N}\frac{\bigl[R_i-R_j+j-i\bigr]_q}{\bigl[j-i\bigr]_q}
=\prod_{1\leq i<j\leq
N}\frac{q^{(R_i-R_j+j-i)/2}-q^{-(R_i-R_j+j-i)/2}}
{q^{(j-i)/2}-q^{-(j-i)/2}}
\end{equation}
with $R_i$ labeling the number of boxes in the $i$-th row of the
Young tableau corresponding to $R$. In~\cite{Aganagic:2004js} a
different but related definition is used whereby the quantum
dimension is replaced by the quantity $S_{00}\,{\rm dim}_q(R)$
with \beq S_{00}(q,N)=\prod_{1\leq i<j\leq N}[j-i]_q \ . \eeq At
finite $N$ the presence of $S_{00}$ is simply a change in the
overall normalization. In the large $N$ limit it will produce the
contribution of constant maps to the topological string amplitude
\cite{Gopakumar:1998ki,Gopakumar:1998ii}, a universal factor that
can be computed separately (see Appendix~\ref{Constmaps}). We will
include this contribution later on when we compare our results
with topological string theory.

In \cite{Aganagic:2004js} the asymptotic expansion in $\frac1N$ of
the $U(N)$ partition function (\ref{Parti}) was constructed
following closely the strategy proposed in
\cite{Gross:1992tu}--\cite{Gross:1993hu} for ordinary
two-dimensional Yang-Mills theory. The sum was restricted to a
subset of representations called ``composite'' large $N$
representations. These are essentially the representations whose
quadratic Casimir invariant has a leading term of order $N$.
Composite representations are formed by taking the tensor product
of a representation $R$ corresponding to a Young diagram with a
finite number of boxes and a representation $\bar{S}$ which is the
complex conjugate of a representation $S$ associated to another
diagram with a finite number of boxes. The resulting large $N$
expansion essentially factorizes into two copies of a simpler
chiral topological string expansion, but with a couple of
important subtleties. Firstly, one should include in the
definition of $Z_{\rm top}$ a sum over a $U(1)$ degree of freedom
identified with a Ramond-Ramond flux through the sphere. Secondly,
and more importantly, the relevant topological string partition
function implies the presence of two stacks of D-branes inserted
in the fibers of $X_p$, represented by two extra sums over
representations with finite numbers of boxes.

The explicit result obtained in \cite{Aganagic:2004js} reads
\begin{equation}
{\cal Z}^q_{\rm YM}=\sum_{l=-\infty}^{\infty}~\sum_{\hat
R_{(1)},\hat R_{(2)}}{\cal Z}^{q{\rm YM},+}_{\hat R_{(1)},\hat
R_{(2)}}(t+p\,g_s l;p)\, {\cal Z}^{q{\rm YM},-}_{\hat R_{(1)},\hat
R_{(2)}}(\,\bar{t}-p\,g_s l;p) \label{AgZexpl}\end{equation} where
the second sum runs through irreducible representations $\hat
R_{(1)},\hat R_{(2)}$ of $SU(N)$ with \beq {\cal Z}^{q{\rm
YM},-}_{\hat R_{(1)},\hat R_{(2)}}(\,\bar{t};p)=(-1)^{|\hat
R_{(1)}|+ |\hat R_{(2)}|}\, {\cal Z}^{q{\rm YM},+}_{\hat
R_{(1)}^\top,\hat R_{(2)}^\top}(\,\bar{t};p) \ .
\label{Zpmrel}\eeq The K\"ahler modulus $t$ parameterizes the area
of the sphere $\PP^1$ and is given by \beq
t=(p-2)\,\frac{N\,g_s}{2}+\ii\theta\label{tKahlermod}\eeq in terms of the
original parameters of the gauge theory (in this paper we set
$\theta=0$). The symbol $|\hat R|$ is the
total number of boxes of the Young tableau of the $SU(N)$
representation $\hat R$. The chiral block ${\cal Z}^{q{\rm
YM},+}_{\hat R_{(1)},\hat R_{(2)}}(t;p)$ agrees exactly with the
perturbative topological string amplitude on $X_p$
\cite{Bryan:2004iq} with two stacks of D-branes inserted in the
fiber. It depends explicitly on the choice of two arbitrary Young
tableaux which correspond to the boundary degrees of freedom of
the fiber D-branes. When all the Young tableaux are taken to be
trivial, i.e. $\hat R_{(1)}=\hat R_{(2)}=0$, one recovers the
expected closed topological string partition function. The chiral
and anti-chiral parts are sewn together along the D-branes and
summed over them.

The full non-chiral partition function also admits a standard description in
terms of toric geometry. As we show explicitly in Section~\ref{toric},
the fibration (\ref{Xfib}) is a toric manifold and the chiral
block ${\cal Z}^{q{\rm YM},+}_{\hat R_{(1)},\hat R_{(2)}}(t;p)$ can be
written in terms of the topological vertex $C_{\hat R_{(1)} \hat
  R_{(2)} \hat R_{(3)}}(q)$~\cite{Aganagic:2003db} as
\bea {\cal Z}^{q{\rm YM},+}_{\hat R_{(1)},\hat
R_{(2)}}(t;p)&=&{\sf Z}_0(q)~ q^{{\kappa_{\hat
R_{(1)}}}/{2}}~\e^{-\frac{t(|\hat R_{(1)}|+| \hat
R_{(2)}|)}{p-2}}\nonumber\\ && \times\, \sum_{\hat R}\e^{-t|\hat
R|}~q^{{(p-1)\kappa_{\hat R_{(1)}}}/{2}}~ C_{0\hat R_{(1)}\hat
R^\top}(q)\,C_{0\hat R\hat R_{(2)}}(q) \ ,
\label{chiraltopvertex}\eea where $\kappa_{\hat R}$ is related to
the Young tableau labels through \beq \kappa_{\hat
R}=\sum_{i=1}^{N-1}\hat R_i\big(\hat R_i-2i+1\big)\eeq and ${\sf
Z}_0(q)$ represents the contribution from constant string maps
(see Appendix~\ref{Constmaps}). This is the partition function of
the topological A-model on $X_p$ with non-compact Lagrangian
D-branes inserted at two of the four lines in the web diagram. The
D-branes are placed at a well-defined ``distance" $t/(p-2)$ from
the sphere, thereby introducing another geometrical parameter.

The extra sum over the integer $l$ originates from the $U(1)$
degrees of freedom contained in the original gauge group $U(N)$
and can be interpreted as a sum over Ramond-Ramond fluxes through
the sphere \cite{Vafa:2004qa}. The sum over the fiber D-branes is
instead related to the fact that the Calabi-Yau manifold $X_p$ is
non-compact and has more moduli coming from the non-compact
directions \cite{Aganagic:2005dh}. Note that the sum over the
``external" branes in the full partition function enters on the
same footing as the sum over the topological string amplitude
constituents. This ``external" sum is weighted with a different
K\"ahler parameter \beq\hat{t}=\frac t{p-2}=\frac{N\,g_s}2\ ,
\label{hattKahlermod}\eeq and the partition function therefore
effectively depends on two parameters. The observation above
suggests that $\hat{t}$ could have an interpretation as a true
K\"ahler modulus. As we will demonstrate in the following, this
follows from a different definition of the chiral gauge theory
which is directly connected to the ordinary Yang-Mills one and
which leads to a closed topological string theory by itself. The
chiral expansion we propose arises from restricting the original
sum to only those representations whose Young diagrams contain a
finite number $n$ of boxes. Coupled representations will not be
considered in this paper.

\subsection{Chiral expansion\label{Chiralexp}}

We now describe the chiral expansion explicitly. The second Casimir
invariant for $U(N)$ representations $R$ has the form \beq
C_2(R)=\kappa_{\hat R}+N\,n-\frac{n^2}{N}+\frac{m^2}{N} \ , \eeq where
$m$ is the $U(1)$ charge
\beq m=n+N\,r\eeq with $r\in\zed$. The row labels of the $U(N)$
representations $R$ are related to those of $SU(N)$ representations
$\hat R$ by $R_i=\hat R_i+r$ for $i=1,\dots,N-1$ and $R_N=r$. The
partition function is a sum over the total number of boxes $n$, the
$SU(N)$ Young tableaux
$\hat{R}$ with $n$ boxes and the $U(1)$ degree of freedom $r$, giving
\beq
{\cal Z}^{q{\rm
YM}}_{\rm chiral}=\sum_{r=-\infty}^{\infty}\e^{-\frac{N\,g_sp}{2}\,
r^2}~\sum_{n=1}^{\infty}~\sum_{\hat
R}\,{\rm dim}_q\big(\hat
R\big)^2~\exp\left[-\frac{N\,g_sp}{2}\,\left(n+\frac{\kappa_{\hat{R}}}{N}
\right)-g_sp\,n\,r\right] \ . \eeq
Let us focus on the $r=0$ sector of vanishing Ramond-Ramond flux
through $S^2$, whose partition function reads \beq {\cal Z}^{q{\rm
YM},0}_{\rm chiral}=\sum_{n=1}^{\infty}~\sum_{\hat R}\,{\rm dim}_q\big(\hat
R\big)^2~\exp\left[-\frac{N\,g_sp}{2}\,\left(n+\frac{\kappa_{\hat{R}}}{N}
\right)\right] \ . \label{ZqYMchiral0}\eeq

To proceed further, we need to understand the structure of the
quantum dimension ${\rm dim}_q(\hat R)$. It can be conveniently
expressed as \cite{Aganagic:2004js} \beq {\rm dim}_q\big(\hat
R\big)=\e^{\frac{g_sN}{2}\,n}\,W_{\hat R}\big(q^{-1}
\big)\,\prod^{c_{\hat
R}}_{i=1}~\prod_{j=1}^{\hat{R}_i}\left(1-q^{j-i}~\e^{-g_sN}\right)
\ , \label{quantdimconv}\eeq where $c_{\hat R}$ is the number of
rows in $\hat R$ and $W_{\hat R}(q)=W_{\hat R0}(q)$ is related to
the large $N$ limit of the modular $S$-matrix of the $SU(N)$ WZW
model and $SU(N)$ Chern-Simons gauge theory by \beq W_{\hat R\hat
Q}(q)=\lim_{N\to\infty}\,q^{\frac{N(|\hat R|+|\hat Q|)}{2}}~
\frac{S_{\hat R\hat Q}(q,N)}{S_{00}(q,N)} \ . \eeq A convenient
way to parameterize the different Young diagrams $\hat R$ with box
number $n$ is as follows. Given a partition \beq
n=n_1+n_2+\dots+n_{i_{\rm max}}\eeq of $n$ with $n_1\geq
n_2\geq\dots\geq n_{i_{\rm max}}$, one obtains the list
$d(n)=(n_1,n_2,\dots,n_{i_{\rm
    max}})$. With $n_i=\hat{R}_i$ the number of boxes in the
$i$-th row of $\hat R$, we have \beq {\rm dim}_q\big(\hat
R\big)=\e^{-\frac{g_sN}{2}\,n}\,W_{\hat
R}\big(q^{-1}\big)\,\prod^{i_{\rm max}}_{i=1}~\prod_{j=1}^{n_i}
\left(1-q^{j-i}~\e^{-g_sN}\right) \ . \eeq By using the relation
\beq W_{\hat R}\big(q^{-1}\big)=(-1)^n\,q^{-{\kappa_{\hat
R}}/{2}}\,W_{\hat R}\big(q \big) \eeq we can rewrite the chiral
partition function as \beq {\cal Z}^{q{\rm YM},0}_{\rm
chiral}=\sum_{n=1}^{\infty}~\sum_{\hat
R}\e^{-\frac{N\,g_s(p-2)}{2}\,n}~\e^{-\frac{g_s(p-2)}{2}\,\kappa_{\hat{R}}}\,
W_{\hat R}(q)^2\,\prod^{i_{\rm max}}_{i=1}~
\prod_{j=1}^{n_i}\left(1-q^{j-i}~\e^{-g_sN}\right)^2 \ .
\label{chiralpartfnrewrite}\eeq The explicit expression
\cite{Aganagic:2002qg} \beq W_{\hat R}(q)=q^{{\kappa_{\hat
R}}/{4}}\, \prod_{1\leq i<j\leq i_{\rm
max}}\frac{[n_i-n_j+j-i]_q}{[j-i]_q}\, \prod_{i=1}^{i_{\rm
max}}~\prod_{k=1}^{n_i}\, \frac{1}{[k-i+i_{\rm max}]_q} \eeq shows
that the general structure of $W_{\hat R}(q)$ is of the form
${q^\alpha}/{\prod_{\beta,\gamma}\,(1-q^\beta)^\gamma}$ for some
$\alpha,\beta,\gamma>0$, and the leading behaviour as $g_s\to 0$
is simply determined by the total number of boxes $n$ as $W_{\hat
R}(q)\simeq g_s^{-n}$.

We are ready now to make the connection with the topological
string expansion. In terms of the K\"ahler modulus
(\ref{tKahlermod}), the chiral partition function is simply
rewritten as \beq {\cal Z}^{q{\rm YM},0}_{\rm
chiral}=\sum_{n=1}^{\infty}\e^{-t\,n}\, \sum_{\hat
R}\e^{-\frac{g_s (p-2)}{2}\,\kappa_{\hat R}}\,W_{\hat
R}(q)^2\,\prod^{i_{\rm max}}_{i=1}~
\prod_{j=1}^{n_i}\left(1-q^{j-i}~\e^{-\frac{2\,t}{p-2}} \right)^2
\ . \label{parto}\eeq The string expansion is obtained by
expanding the partition function in powers of the string coupling
constant $g_s$ with $t$ fixed. Notice that this is not a power
series in $\e^{-t}$ at any given order in $g_s$, as one would
expect in a conventional topological string perturbative expansion
with a single K\"ahler modulus. There is also a power series in
$\e^{-\frac{2\,t}{p-2}}$ which in the coupled partition function
appears as fiber D-brane contributions whose distance from the
$S^2$ is parameterized by the K\"ahler modulus
(\ref{hattKahlermod}). We will come back to this point later on.
The general structure of the free energy ${\cal F}_{\rm
chiral}^{q{\rm YM}}=\log {\cal Z}^{q{\rm
    YM},0}_{\rm chiral}$ is given by
\beq {\cal F}_{\rm chiral}^{q{\rm YM}}
=\sum_{g=0}^\infty\,g_s^{2g-2}~{\cal F}_g
\eeq
with
\beq
{\cal F}_g\big(t\,,\,\hat t\,;\,p\big)=
\sum_{n=1}^{\infty}\e^{-n\,t}\,\sum_{k=0}^{2n}\e^{-2\,k\,\hat t}~
{\rm N}^{g}{}^{~}_{n,k}(p) \ ,\label{freen}\eeq
where $n$ labels the winding numbers of holomorphic maps from genus
$g$ Riemann surfaces into the local threefold $X_p$ and ${\rm
  N}^{g}{}^{~}_{n,k}(p)\in\rat$ are generalized Gromov-Witten
invariants. The conventional Gromov-Witten invariants ${\rm
  N}^{g}{}^{~}_{n}={\rm N}^{g}{}^{~}_{n,k=0}$ of the local Calabi-Yau
threefold $X_p$ ``count'' the genus~$g$ worldsheet instantons
of degree~$n$ in $X_p$ and arise when the fiber D-branes are
ignored. The geometrical meaning of the generalized invariants ${\rm
  N}^{g}{}^{~}_{n,k}(p)$ for $k>0$ will be elucidated later on.

Let us specialize at this point to the genus zero contribution, as it
is this term which will exhibit the interesting phase structure of the
theory. The higher genus contributions are similarly dealt with (The
genus one free energy is worked out in
Appendix~\ref{Genus1partfn}). By parameterizing the free energy as
\beq {\cal F}_0\big(t\,,\,\hat t\,;\,p\big)=\left(1-\e^{-2\,\hat t}\, \right)
^2~\sum_{n=1}^\infty \e^{-n\,t} ~F_n\big(\hat t\,;\,p\big) \ , \eeq
we find that the first six
contributions are given by \bea
F_1&=&1 \ , \nonumber\\[5pt]
F_2&=&\frac{\e^{-4\,\hat t}}
  {8}\,\left[1 + 4\,p + 2\,p^2 + \e^{2\,\hat t}\,\left( 2 - 4\,p^2
    \right)+\e^{4\,\hat t}\,\left( 1 - 4\,p + 2\,p^2 \right)
  \right] \ , \nonumber\\[5pt]
F_3&=&\frac{\e^{-8\,\hat t}}{54}\,\left[2 + 18\,p + 45\,p^2 +
36\,p^3 + 9\,p^4 + 6~\e^{4\,\hat t}\,\left( 1 - 9\,p^2 +
9\,p^4 \right)\right.\nonumber\\
&&+  \left.  ~\e^{8\,\hat t}\,\left( 2 - 18\,p + 45\,p^2 -
36\,p^3 + 9\,p^4 \right)+
    2~\e^{6\,\hat t}\,\left(2 - 9\,p - 9\,p^2 + 36\,p^3 - 18\,p^4
\right)\right.\nonumber\\ &&+\left.
    2~\e^{2\,\hat t}\,\left(2 + 9\,p - 9\,p^2 - 36\,p^3 - 18\,p^4
\right)\right] \ , \nonumber\\[5pt]
F_4&=&\frac{\e^{-12\,\hat t}}{192}\,\Bigl[3 + 44\,p + 214\,p^2 + 448\,
p^3 + 432\,p^4 + 192\,p^5 + 32\,p^6\Bigr.\nonumber\\
&&\left.+~
    \e^{10\,\hat t}\,\left( 6 - 56\,p + 60\,p^2 + 416\,p^3 - 1008\,
p^4 + 768\,p^5 - 192\,p^6 \right)\right.\nonumber\\
&&\left.+~
    \e^{12\,\hat t}\,\left( 3 - 44\,p + 214\,p^2 - 448\,p^3 + 432\,
p^4 - 192\,p^5 + 32\,p^6 \right)\right.\nonumber\\
&&\left.+\,
    2~\e^{2\,\hat t}\,\left(3 + 28\,p + 30\,p^2 - 208\,p^3 - 504\,
p^4 - 384\,p^5 - 96\,p^6 \right)\right.\nonumber\\
&&\left.+\,
    4~\e^{6\,\hat t}\,\left(3 - 62\,p^2 + 216\,p^4 - 160\,p^6
\right)\right.\nonumber\\ &&\left.+~
    \e^{8\,\hat t}\,\left( 9 - 44\,p - 150\,p^2 + 512\,p^3 + 144\,
p^4 - 960\,p^5 + 480\,p^6 \right)\right.\nonumber\\
&&\left.+~
    \e^{4\,\hat t}\,\left( 9 + 44\,p - 150\,p^2 - 512\,p^3 + 144\,
p^4 + 960\,p^5 + 480\,p^6 \right) \right] \ , \nonumber\\[5pt]
F_5&=& \nonumber\frac{\e^{-16\,\hat t}}{3000}\, \Bigl[  24 +
500\,p + 3750\,p^2 + 13500\,p^3 + 25875\,p^4 + 27500\,p^5 +
16250\,p^6 + 5000\,p^7\Bigr.\\ && \nonumber+\, 625\,p^8
      +\e^{14\,\hat t}\,\big( 48 - 700\,p + 2250\,p^2 +
       5000\,p^3 - 38750\,p^4 + 76700\,p^5-\nonumber 69500\,p^6
\\ && \nonumber+\, 30000\,p^7 - 5000\,p^8 \big)  +
     \e^{16\,\hat t}\,\big( 24 - 500\,p + 3750\,p^2 - 13500\,p^3
+ 25875\,p^4
     \\ &&\nonumber-\, 27500\,p^5 + 16250\,p^6 - 5000\,p^7 + 625\,p^8
\big)  -2~\e^{2\,\hat t}\,\big( -24 - 350\,p -
      1125\,p^2\\ && \nonumber
+\, 2500\,p^3 + 19375\,p^4 + 38350\,p^5 + 34750\,p^6 + 15000\,p^7 +
2500\,p^8 \big) -      10~\e^{8\,\hat t}\,\big( 12\\
&& \nonumber-\, 450\,p^2 + 3325\,p^4 - 7250\,p^6 + 4375\,p^8 \big) +
4~\e^{12\,\hat t}\,\big( 18  - 175\,p - 125\,p^2 +
3500\,p^3\\ && \nonumber-\, 5125\,p^4 - 7950\,p^5 + 23000\,p^6 -
17500\,p^7 + 4375\,p^8 \big)  +      4~\e^{4\,\hat t}\,
\big( 18 + 175\,p \\ &&\nonumber -\, 125\,p^2 -
3500\,p^3 - 5125\,p^4 + 7950\,p^5 + 23000\,p^6 + 17500\,p^7 +
4375\,p^8 \big)\\ &&\nonumber+\,
      2~\e^{10\,\hat t}\,\big(48 - 250\,p - 1625\,p^2 + 5500\,
p^3 + 8375\,p^4 - 28250\,p^5 - 1250\,p^6\\ &&
\nonumber+\, 35000\,p^7 - 17500\,p^8 \big)+
      2~\e^{6\,\hat t}\,\big(48 + 250\,p - 1625\,p^2
- 5500\,p^3\\ &&  +\,\Bigl. 8375\,p^4 + 28250\,p^5 - 1250\,p^6 -
35000\,p^7 - 17500\,p^8 \big)  \Bigr]  \ , \nonumber\\[5pt]
 F_6&=& \nonumber\frac{\e^{-20\,\hat t}}{2160} \,\Bigl[ 10 + 274\,p +
2837\,p^2 + 14940\,p^3 + 44955\,p^4 + 81756\,p^5 + 92250\,p^6 +
64800\,p^7\Bigr.\\ && \nonumber+ \,27540\,p^8+6480\,p^9+648\,p^{10}
      +\e^{18\,\hat t }\,\big( 20 - 404\,p + 2230\,p^2 +
       630\,p^3 - 44880\,p^4 \\ && \nonumber+\,170154\,p^5 -\nonumber
       304530\,p^6 +303120\,p^7 - 171720\,p^8
       +51840\,p^9-6480\,p^{10} \big)\\ && \nonumber+~
     \e^{20\,\hat t}\,\big( 10 - 274\,p + 2837\,p^2 -14940 \,p^3 +44955\,p^4 -
    81756\,p^5 +92250\,p^6 -64800\,p^7\\ &&\nonumber +\,27540
    \,p^8-6480\,p^9+648\,p^{10}  \big)+
      2~\e^{2\,\hat t}\,\big(10 + 202\,p +
      1115\,p^2
- 315\,p^3\\ && \nonumber -\, 22440\,p^4 - 85077\,p^5 - 152265\,p^6
- 151560\,p^7 - 85860\,p^8 -25920\,p^9+3240\,p^{10} \big) \\ &&
\nonumber+\,
 3~\e^{ 16\,\hat t}\,\big( 10 - 148\,p+225\,p^2 + 3780\,p^3 - 15035
 \,p^4 +3168\,p^5+72610\,p^6-152640\,p^7 \\ && \nonumber+\,
136620\,p^8-58320\,p^9+9720\,p^{10}  \big) +3~\e^{4\,\hat t}\,\big(
10 +148\,p+225\,p^2 - 3780\,p^3 - 15035 \,p^4 \\ &&
\nonumber-\,3168\,p^5+72610\,p^6+152640\,p^7+
136620\,p^8+58320\,p^9+9720\,p^{10}  \big) \\ && \nonumber +\,
12~\e^{10\,\hat t}\,\big(5-297\,p^2 + 3760\,p^4 - 16325\,p^6 +
26460\,p^8 -13608\,p^{10} \big)  \\ &&\nonumber +\,4~\e^{14\,\hat
t}\,\big(10 + 101\,p-
 285\,p^2 +
3555\,p^3 - 1860\,p^4 - 26109\,p^5 +42780\,p^6 \\ &&\nonumber+\,
24120\,p^7 - 100440\,p^8-77760\,p^9-19440\,p^{10} \big)  +
      4~\e^{ 6\,\hat t}\,\big(10 +101\,p - 285 \,p^2 \\ &&\nonumber-\,
3555\,p^3 -1860 \,p^4 +26109\,p^5 +42780\,p^6 -24120\,p^7 -
100440\,p^8-77760\,p^9\\ && \nonumber-\,19440\,p^{10}\big)+
     2~\e^{12\,\hat t}\,\big( 25 -137\,p -1410\,p^2
+4860\,p^3+ 14955\,p^4 -45738\,p^5\\ && \nonumber-\, 39360\,p^6 +
146160\,p^7-11340\,p^8 - 136080\,p^9 + 68040\,p^{10} \big) +
2~\e^{8\,\hat t}\,\big( 25 \\ && \nonumber+\,137\,p -1410\,p^2
-4860\,p^3+ 14955\,p^4 +45738\,p^5- 39360\,p^6 \\ &&  -\Bigl.
146160\,p^7-11340\,p^8 + 136080\,p^9 + 68040\,p^{10} \big)
  \Bigr] \ . \label{frees}\eea
Note that ${\rm N}^{0}{}^{~}_{n,k}(p)$ is a polynomial of degree
$2n-2$ in $p$.

As a first application of these results, let us check the
consistency of our computations with ordinary large $N$ Yang-Mills
theory. It is possible to find a limit in which the undeformed
theory is recovered
\cite{Caporaso:2005ta,Arsiwalla:2005jb,Jafferis:2005jd}, namely $p\to
\infty$ with $A=2 \,p \,\hat  t=p\,N\,g_s$ kept fixed. The free
energies (\ref{frees}) in this limit should reproduce the
analogous quantities obtained in ordinary chiral Yang-Mills theory
from the Gross-Taylor string expansion. By explicitly performing
this limit we find \beq \Phi_{0}(A):=\lim_{\stackrel{ \scriptstyle
p\to\infty}{\scriptstyle A=2\,p\,\hat t}}\, {\cal
F}_0\big(t\,,\,\hat t\,;\,p\big)=\sum_{n=1}^\infty \e^{-n\,
A/2}~\phi_n(A) \eeq with \bea
\phi_1&=&1 \ , \nonumber\\[5pt]
\phi_2&=&\mbox{$\frac{1}{2} - A + \frac{1}{4}\,A^2$} \ , \nonumber\\[5pt]
\phi_3&=&\mbox{$\frac{1}{3} - 2\,A + 3\,A^2 - \frac{4}{3}\,A^3 +
\frac{1}{6}$}\,A^4 \ , \nonumber\\[5pt]
\phi_4&=&\mbox{$\frac{1}{4} - 3\,A + \frac{21}{2}\,A^2 -
\frac{43}{3}\,A^3 + \frac{33}{4}\,A^4 - 2\,A^5 + \frac{1}{6}\,A^6$}
\ , \nonumber\\[5pt]
\phi_5&=&\mbox{$\frac{1}{5} - 4\,A + 25\,A^2 - \frac{202}{3}\,A^3
+ \frac{529}{6}\,A^4 - \frac{883}{15}\,A^5 + \frac{121}{6}\,A^6 -
\frac{10}{3}\,A^7 + \frac{5}{24}\,A^8$} \ ,
\nonumber\\[5pt]
\nonumber \phi_6&=&\mbox{$\frac{1}{6} - 5\,A + \frac{195 }{4}\,A^2
- \frac{647}{3}\,A^3 + \frac{1489}{3}\,A^4 - \frac{3178}{5}\,A^5 +
\frac{1871}{4}\,A^6 - \frac{598}{3}\,A^7$}
\\ &&
+\,\mbox{$
48\,A^8-6\,A^9+\frac{3}{10}\,A^{10}$} \ . \label{hurwitz}\eea We find
exact agreement at this order with the results obtained in
\cite{Taylor:1994zm,Crescimanno:1994eg} for the ordinary chiral
QCD$_2$ string.

The second application is somewhat more sophisticated and will be
discussed in Sections~\ref{GTexp} and~\ref{stringseries}. From
(\ref{frees}) it is possible to compute the generalized
Gromov-Witten invariants of the topological string theory on
$X_p$. Because ordinary chiral QCD$_2$ computes Hurwitz numbers
which are encoded in the coefficients of the area polynomials in
(\ref{hurwitz})\cite{Taylor:1994zm}, our limiting procedure
establishes a direct link between the combinatorics of branched
covering maps to $\PP^1$ and the geometry of rational curves
embedded in Calabi-Yau threefolds. In Section~\ref{stringseries}
we will use the information encoded in these free energies to
estimate the radius of convergence of the string perturbation
series and to describe the phase transitions in the chiral gauge
theory.

\subsection{Toric geometry and the topological vertex\label{toric}}

We will now derive the toric description of the local Calabi-Yau
threefold (\ref{Xfib}), and by using the formalism of the
topological vertex \cite{Marino:2004uf,Aganagic:2003db} we will
find the toric geometry associated to the large $N$ limit of the
chiral $q$-deformed Yang-Mills theory. The total space $X_p$ may
be regarded as a special Lagrangian $\mathbb{T}^2 \times
\mathbb{R}$ fibration over $\mathbb{R}^3$. With $(z_i)_{i=1}^4$
coordinates on the complex linear space $\complex^4$, consider the
real equation \bea \label{C4realeq}
\mu_p:=|z_1|^2+|z_4|^2-p|z_2|^2+(p-2)|z_3|^2=t
 \eea
and the $U(1)$ group action on $\complex^4$ given by \bea \label{U1actionC4}
\big(z_1\,,\,z_4\,,\,z_2\,,\,z_3
\big)~\longmapsto ~\big(\e^{\ii \alpha}\,z_1\,,\,\e^{\ii
\alpha}\,z_4\,,\,\e^{-\ii p\, \alpha}\,z_2\,,\,\e^{\ii
(p-2)\alpha}\,z_3\big) \ .
 \eea
Then $X_p=\mu_p^{-1}(t)/U(1)$. For $z_2=z_3=0$ eq.~(\ref{C4realeq})
describes a sphere whose area is proportional to $t$. Thus $(z_1,z_4)$
can be taken as homogeneous coordinates for the base $\PP^1$ of the
fibration (\ref{Xfib}), while $(z_2,z_3)$ may be regarded as
coordinates of the fibers.

This realization defines $X_p$ as a symplectic quotient. For this,
we regard $X_p$ as a union of local coordinate patches each
symplectomorphic to
$\mathbb{C}^3$~\cite{Aganagic:2003db,Marino:2004uf}. There are two
patches because the base $\mathbb{P}^1$ is given by the equation
$|z_1|^2+|z_4|^2=t$ with one of $z_1$ or $z_4$ non-zero. In each
patch we write down moment maps $(r_\alpha,r_\beta,r_\gamma)$
whose image gives global coordinates for the base $\real^3$ and
which generate three Hamiltonian flows on $\complex^3$ with
respect to its standard symplectic structure. The torus fiber
$\torus^2$ corresponds to the circle actions generated by
$r_\alpha$ and $r_\beta$, while $r_\gamma$ generates the real line
$\real$. The local $\complex^3$ geometry of $X_p$ is represented
by an oriented trivalent planar graph which encodes the
degeneration locus of the fibration in the base $\real^3$, drawn
in the $r_\gamma=0$ plane. An edge of the graph is labelled by an
integer vector $(n,m)\in\zed^2$ that corresponds to the generator
of the homology group $H_1(\torus^2,\zed)$ which is the shrinking
cycle. Changing the edge orientations sends $(n,m)\mapsto(-n,-m)$
and does not alter the Calabi-Yau geometry. The local geometry is
described by assigning to each patch three integer vectors $\vec
v_a=(n_a,m_a)$, $a=1,2,3$ which single out the degenerating cycles
unambiguously up to $SL(2,\zed)$ modular transformations of
$\torus^2$. The lines in the base $\real^3$ where the $\torus^2$
fibers degenerate are correlated with the zeroes of the
corresponding moment maps. The Calabi-Yau condition is encoded in
the requirement $\sum_a\,\vec v_a=(0,0)$. The conditions $|\vec
v_a\wedge\vec v_b|=1$, $a<b$ ensure smoothness of $X_p$, where
$\wedge$ denotes the symplectic product on the vector space
$H_1(\torus^2,\real)$.

Let us now explicitly display the two $\complex^3$ patches of the
Calabi-Yau space $X_p$.

\medskip

\begin{figure}[hbt]
\begin{center}
\epsfxsize=1.5 in\epsfbox{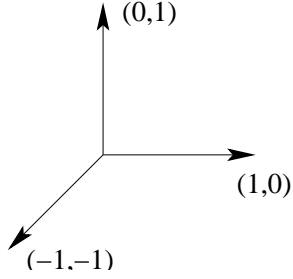}
\end{center}
\caption{Toric graph for the patch $U_4(z_1,z_2,z_3)$ of $X_p$
  representing the singular locus in the base $\real^3$ with
  global coordinates $(r_\alpha,r_\beta,r_\gamma)$.}
\label{U4patch}\end{figure}

\noindent $\mbf{z_4\neq0}$: \  In this case we can use
(\ref{C4realeq}) to solve for the modulus of $z_4$ in terms of
$z_1,z_2,z_3$ and gauge away its phase by dividing by the $U(1)$
action (\ref{U1actionC4}) of the symplectic quotient construction.
This defines the patch $U_4(z_1,z_2,z_3)\cong\complex^3$. The
Hamiltonians which generate the homology cycles of the $\torus^2$
fiber are defined by
\bea \nonumber r_\alpha&=&|z_2|^2-|z_1|^2 \ , \\
r_\beta&=&|z_3|^2-|z_1|^2
 \eea
with the torus action \bea \e^{\ii \alpha\,r_\alpha+
\ii \beta\, r_\beta}\,:\,
\big(z_1\,,\,z_2\,,\,z_3\big) ~\longmapsto~\big(\e^{-\ii
(\alpha+\beta)}\,z_1\,,\,\e^{\ii
\alpha}\,z_2\,,\,\e^{\ii \beta}\,z_3\big) \ .
 \eea
The $\real$ fiber is generated by $r_\gamma={\rm Im}(z_1z_2z_3z_4)$. The
degeneration locus corresponds to zero sets of these moment
maps. Using (\ref{C4realeq}) one finds that the $(1,0)$ cycle
generated by $r_\beta$ degenerates over the line
$z_1=z_3=0$ where $r_\beta=r_\gamma=0$ and $r_\alpha\geq0$, the
$(0,1)$ cycle generated by $r_\alpha$ degenerates over $z_1=z_2=0$
where $r_\alpha=r_\gamma=0$ and $0\leq r_\beta<t$, and the $(-1,-1)$ cycle
generated by $-(r_\alpha+r_\beta)$ degenerates over $z_2=z_3=0$ where
$r_\alpha-r_\beta=r_\gamma=0$ and $-t<r_\alpha\leq0$. Imposing the
Calabi-Yau and smoothness conditions, and defining $\vec v_1=(-1,-1)$,
we arrive at the basis for the degeneration locus in
$H_1(\torus^2,\zed)$ given by
\bea \vec v_1=
(-1,-1) \ ,\ \ \ \ \vec v_2= (0,1) \ ,\ \ \ \ \vec v_3= (1,0) \ .
\label{canframing}\eea
The graph representing this locus is depicted in Fig.~\ref{U4patch}.

\medskip

\begin{figure}[hbt]
\begin{center}
\epsfxsize=1.5 in \epsfbox{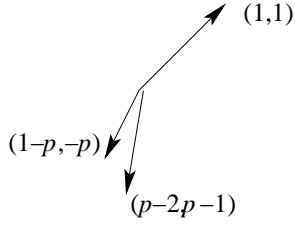}
\end{center}
\caption{Toric graph for the patch $U_1(z_2,z_3,z_4)$
  of $X_p$ representing the singular locus in the base $\real^3$ with
  global coordinates $(r_\alpha,r_\beta,r_\gamma)$.}
\label{U1patch}\end{figure}

\noindent $\mbf{z_1 \neq 0}$: \ In this case we can solve for
$z_1$ and produce the patch $U_1(z_2,z_3,z_4)\cong\complex^3$. The
same Hamiltonians as before generate the $\torus^2\times\real$
fiber, except that now $z_1$ is no longer a natural coordinate for
the patch and so we use (\ref{C4realeq}) to write \bea \nonumber
r_\alpha&=&|z_2|^2-|z_1|^2=-t+(1-p)|z_2|^2+(p-2)|z_3|^2
+|z_4|^2 \ , \\
r_\beta&=&|z_3|^2-|z_1|^2=-t-p|z_2|^2+(p-1)|z_3|^2+|z_4|^2
 \eea
with the torus action \bea \e^{\ii \alpha \,r_\alpha+
\ii \beta \,r_\beta}\,:\,
\big(z_2\,,\,z_3\,,\,z_4\big)
~\longmapsto~\big(\e^{\ii(1-p)\alpha-
\ii p\,\beta}\,z_2\,,\,\e^{\ii
(p-2)\alpha+\ii(p-1)\beta}\,z_3\,,\,\e^{\ii(\alpha+ \beta)}\,
z_4\big) \ . \nonumber\\
\eea
Again by using (\ref{C4realeq}) one finds that the $(p-2,p-1)$ cycle
generated by $(p-2)r_\alpha+(p-1)r_\beta$ degenerates over the line
$z_2=z_4=0$ where $(p-1)r_\alpha+(p-2)r_\beta=(3-2p)t$, $r_\gamma=0$
and $-t\leq r_\alpha<0$, the $(1-p,-p)$ cycle degenerates for
$z_3=z_4=0$ where $p\,r_\alpha+(p-1)r_\beta=(1-2p)t$, $r_\gamma=0$ and
$r_\alpha\leq t$, and finally the $(1,1)$ cycle generated by
$r_\alpha+r_\beta$ degenerates over $z_2=z_3=0$ where
$r_\alpha-r_\beta=r_\gamma=0$ and $-t\leq r_\alpha<0$. Defining $\vec
v_1^{\,\prime} = (1,1)$ thereby gives the degeneration basis
\bea \vec v_1^{\,\prime}= (1,1) \ , \ \
\ \ \ \ \vec v_2^{\,\prime}=(1-p,-p) \ , \ \ \ \ \ \
\vec v_3^{\,\prime}= (p-2,p-1)
\eea
depicted in Fig.~\ref{U1patch}.

\begin{figure}[hbt]
\begin{center}
\epsfxsize=2.0 in \epsfbox{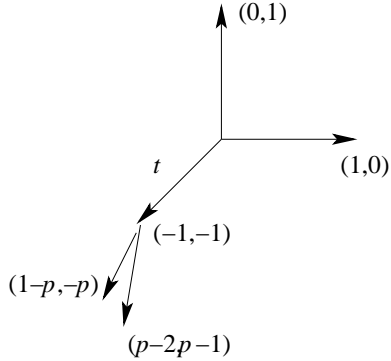}
\end{center}
\caption{Toric diagram of $X_p={\cal O}(-p)\oplus{\cal
    O}(p-2)\longrightarrow\PP^1$. The manifold is built by gluing its two
  $\complex^3$ patches together along their common
but oppositely oriented sphere $\PP^1$ with K\"ahler modulus~$t$.}
\label{toricX}\end{figure}

Note that both patches share the common edge where $z_2=z_3=0$ through
the orientation reversing symmetry $(-1,-1)\leftrightarrow(1,1)$ of their
graphs. The length of this edge is the K\"ahler parameter $t$ and it
represents the base $\PP^1$ of the fibration (\ref{Xfib}). The
threefold $X_p$ is finally obtained by gluing the two $\complex^3$
patches together along this common edge. The transition functions are
given by $SL(2,\zed)$ modular transformations of the $\torus^2$ fibers
between the patches. The graph encoding the
toric geometry of $X_p$ is depicted in Fig.~\ref{toricX}.

We can now construct generic topological string amplitudes on $X_p$ by
using the formalism of the topological vertex~\cite{Aganagic:2003db}
(see~\cite{Marino:2004uf} for a review). The toric
geometry of $X_p$ is encoded in a planar graph obtained by gluing
trivalent vertices representing the $\mathbb{C}^3$ patches. The basic
object associated to the trivalent vertices is the open topological
string vacuum amplitude $C_{\hat R_{(1)}\hat R_{(2)}\hat R_{(3)}}(q)$
  on the trivial $\complex^3$ geometry, where $\hat R_{(a)}$, $a=1,2,3$ are
  $SU(N)$ representation labels on the three edges $\vec v_a$ of the graph. This
  defines the cubic topological vertex~\cite{Aganagic:2003db}. It is
  proportional to the combinatorial quantity $\sum_{\cal Y}\,q^{|{\cal
      Y}|}$, where the sum runs over plane partitions $\cal Y$ whose
  edges in the three directions correspond to Young tableaux with the
  shapes $\hat R_{(a)}$~\cite{ORV1}. Explicitly, the topological
  vertex amplitude in the canonical framing (\ref{canframing}) is given by
\bea C_{\hat R_{(1)}\hat R_{(2)}\hat
R_{(3)}}(q)&=&q^{(\kappa_{\hat R_{(2)}}+ \kappa_{\hat
R_{(3)}})/2}~\sum_{\hat Q}~\sum_{\hat Q_{(1)},\hat Q_{(3)}}
N_{\hat Q\hat Q_{(1)}}^{~~~~~~\hat R_{(1)}}\, N_{\hat Q\hat
Q_{(3)}}^{~~~~~~\hat R_{(1)}^\top}\nonumber\\ && \qquad\qquad
\times\, \frac{W_{\hat R^\top_{(2)}\hat Q_{(1)}}(q)\, W_{\hat
R_{(2)}\hat Q_{(3)}}(q)}{W_{\hat R_{(2)}}(q)}
\label{topvertexexpl}\eea where $N_{\hat Q\hat R}^{~~~~\hat S}$
are the $SU(N)$ fusion numbers. It is natural to expect that
closed string amplitudes associated to the toric diagram in
Fig.~\ref{toricX} can be computed by gluing the open topological
string amplitudes associated to the trivalent vertices, in much
the same way that one computes amplitudes in perturbative quantum
field theory by gluing vertices through propagators.

The gluing rules for the topological vertex are quite
simple~\cite{Aganagic:2003db}. First of all, we need to reverse the
orientation of the edge $\vec v_1$, which in the open topological
string amplitude induces the transformation
\beq
C_{\hat R_{(1)}\hat R_{(2)}\hat R_{(3)}}(q)~\longmapsto~
(-1)^{|\hat R_{(1)}|}\,C_{\hat R_{(1)}^\top\hat R_{(2)}\hat
  R_{(3)}}(q)
\eeq
corresponding to the gluing of topological branes to antibranes. Then
we have to take care of the fact that the patch $U_1$ is not in the
canonical framing given by the basis
(\ref{canframing})~\cite{Aganagic:2003db,Marino:2004uf}. This
implies the presence of an additional factor $
(-1)^{- n_1  |\hat R_{(1)}|}\,q^{n_1 \kappa_{\hat R_{(1)}}/2}$ in the
amplitude, where $n_1=|\vec v_3^{\,\prime}\wedge\vec v_3|=p-1$. We glue
together the two vertices with the Schwinger propagator $\e^{-|\hat
  R_{(1)}|t}~\delta_{\hat R_{(1)}',\hat R_{(1)}^\top}$ coming from the
worldsheet instanton action on the lines which represent spheres
$\PP^1$. Collecting all factors, we arrive at the topological
string partition function given by \beq Z^{~}_{{\hat R_{(2)},\hat
R_{(3)}}\atop {\hat R_{(2)}',\hat R_{(3)}'}}(t;p)= \sum_{\hat
R_{(1)}}\e^{-|\hat R_{(1)}|t }\,(-1)^{p|\hat R_{(1)}|}\,
q^{(p-1)\kappa_{\hat R_{(1)}}/2}\,C_{\hat R_{(1)}\hat R_{(2)} \hat
R_{(3)}}(q)\, C_{\hat R_{(1)}^\top\hat R_{(2)}'\hat R_{(3)}'}(q) \
.
 \eeq
This is a generalization of the closed topological string vacuum
amplitude on $X_p$, with representations $\hat R_{(2)},{\hat R}_{(2)}',\hat
R_{(3)},{\hat R}_{(3)}'$ placed on the external legs of the toric
diagram. These representations describe D-brane degrees of freedom
\cite{Aganagic:2003db} corresponding to non-compact special Lagrangian
submanifolds with the topology of $\complex\times S^1$ in the
edges that go to infinity in the toric diagram for $X_p$.

Let us now analytically continue $t=\ii \pi \,p + t'$. As explained in
\cite{Aganagic:2004js,Aganagic:2005dh} there are only two stacks
 of D-branes inserted in the fibers of $X_p$ that correspond to
extra closed string moduli coming from infinity. In the other two
directions, where the D4-branes are wrapped, we
have to consider trivial representations (heuristically this line
bundle should be understood as a degenerate limit of a compact
cycle). These are the cycles $\vec v_3$ and $\vec v^{\,\prime}_2$ and
the condition requires setting $\hat R_3=\hat R_2'=0$. By using the identity
\beq W_{\hat R\hat S^\top}(q)=q^{-\kappa_{\hat S}/2}~C_{0\hat
R\hat S}(q) \label{WC0RSid}\eeq along with the cyclicity of the
topological vertex in its representation labels, we arrive at \beq
Z^{~}_{{\hat R_{(2)},~0~~}\atop {~~0~,\hat
R_{(3)}'}}(t';p)=q^{\kappa_{\hat R_{(2)}}/2}~ \sum_{\hat
R_{(1)}}\e^{-|\hat R_{(1)}|t'}\, q^{(p-2)\kappa_{\hat
R_{(1)}}/2}\,W_{ \hat R_{(3)}'\hat R_{(1)}}(q)\, W_{\hat
R_{(1)}\hat R_{(2)}^\top}(q) \ . \eeq Up to overall normalization
this expression coincides with the chiral partition function
${\cal Z}^{q{\rm YM},+}_{\hat R_{(3)}',\hat R_{(2)}}(t'\,)$
computed in~\cite{Aganagic:2004js} (see
eq.~(\ref{chiraltopvertex})). In particular, the closed
topological string partition function on $X_p$ is given by \beq
Z^{~}_{{0,0}\atop{0,0}}(t';p)= \sum_{\hat R_{(1)}}\e^{-|\hat
R_{(1)}|t'}\, q^{(p-2) \kappa_{\hat R_{(1)}}/2}\, W_{\hat
R_{(1)}}(q)^2 \ . \label{Z0000}\eeq This expression coincides
exactly with the perturbative part of the topological string
amplitude on $X_p$ at the value of the K\"ahler parameter fixed by
the attractor mechanism~\cite{Aganagic:2004js}. It is
straightforward to check that the $k=0$ sector of
eq.~(\ref{parto}) also reproduces the topological string partition
function on $X_p$. The relevant Gromov-Witten invariants are given
in this case by ${\rm
  N}^g{}^{~}_{n}(p)={\rm N}^{g}{}^{~}_{n,0}(p)$. We will now explore this
relationship in more detail.

\subsection{Chiral partition function as a topological string
amplitude\label{chiraltopstring}}

In Section~\ref{Chiralexp} we found that the  chiral partition
function, obtained by disregarding coupled representations, has a
free energy which is nicely organized as a double series expansion
in the parameters $t$ and $\hat t$ according to eq.~(\ref{freen}).
This strongly suggests that $\hat t$ should be regarded as another
K\"ahler parameter. We will now show that this is indeed the case
and also make contact with the amplitude (\ref{Z0000}) which can
be interpreted as the topological string partition function on
$X_p$. For this, let us begin with the simple identity
\beq
\prod^{i_{\rm max}}_{i=1}~\prod_{j=1}^{n_i}\left(1-q^{j-i}~
\e^{-g_sN}\right)=\exp\Bigl[~\sum_{n=1}^{\infty} \frac{1}{n}
\,f_{\hat R}(q^n)~\e^{-n\,g_sN}\,\Bigr] \label{Agid1}\eeq where
\beq f_{\hat R}(q)=\sum_{i=1}^{i_{\rm
max}}~\sum_{j=1}^{n_i}\,q^{j-i} \ . \label{fRqdef}\eeq This can in
turn be written in terms of the function \beq N_{\hat
R}\big(q\,,\,\e^{-g_sN}\big):=\sum_{\hat S}\,(-1)^{|\hat S|}~
\e^{-|\hat S|g_sN}\,W_{\hat S\hat R}(q)\,W_{\hat S^\top}(q)
\label{NhatRdef}\eeq by means of the
identity~\cite{Aganagic:2004js} \beq N_{\hat
R}\big(q\,,\,\e^{-g_sN}\big)= N_{0}(q,N)\,  W_{\hat R}(q)
\,\exp\Bigl[~\sum_{n=1}^{\infty} \frac{1}{n} \,f_{\hat
R}(q^n)~\e^{-n\,g_sN}\,\Bigr] \ . \label{Agids}\eeq

Along with (\ref{WC0RSid}) these identities enable us to rewrite
our chiral partition function (\ref{chiralpartfnrewrite}) as \bea
{\cal Z}_{\rm chiral}^{q{\rm YM},0}&=&{\sf Z}_0(q)\,\sum_{\hat R}~
\sum_{\hat R_{(1)},\hat R_{(2)}}\e^{-t |\hat R|-t_1 |\hat
R_{(1)}|-t_2 |\hat R_{(2)}|}\, (-1)^{|\hat R_{(1)}|+|\hat
R_{(2)}|+p |\hat R|}\,q^{(p-1)\kappa_{\hat R}/2} \nonumber\\ &&
\qquad\qquad \times\, C_{0\hat R_{(1)}\hat R^\top}(q)\,C_{0\hat
R\hat R_{(2)}}(q)\, C_{00\hat R_{(1)}^\top}(q)\, C_{00\hat
R_{(2)}^\top}(q) \ , \label{top3}\eea where $t_1=t_2=2\,\hat t$.
In this computation we have restored the normalization factor
$S_{00}$ and defined (see Appendix~\ref{Constmaps}) \beq {\sf
Z}_0(q)=\lim_{N\to\infty}\,\frac{S_{00}(q,N)}{N_{0}(q,N)}=
M(q)\,\eta(q)^N\,q^{-{N}/{24}} \ , \label{Z0qdef}\eeq where $M(q)$
is the McMahon function and $\eta(q)$ is the Dedekind function.
The factor (\ref{Z0qdef}) accounts for the contribution of
constant string maps into $X_p$ (having winding number $n=0$) and
for quadratic (in $t$) ambiguities in the genus zero partition
function \cite{Aganagic:2004js}, plus some non-perturbative
corrections. The remaining part of (\ref{top3}) represents instead
the full perturbative contribution to a closed topological string
amplitude.

\begin{figure}[hbt]
\begin{center}
\epsfxsize=2.0 in\epsfbox{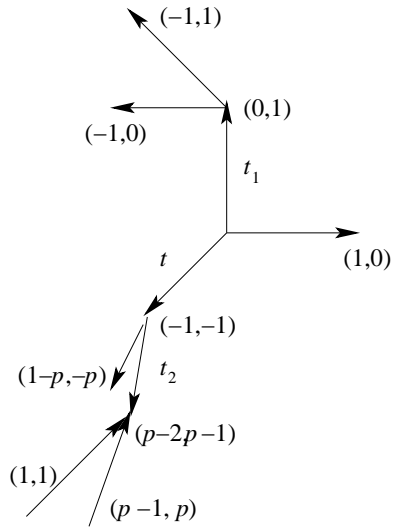}
\end{center}
\caption{Toric diagram describing the topological string expansion of
  the chiral $q$-deformed gauge theory partition function ${\cal Z}_{\rm
    chiral}^{q{\rm YM},0}$.}
\label{toricchiral}\end{figure}

The relevant toric diagram can be read off directly from
eq.~(\ref{top3}) by reversing the topological vertex gluing rules. It
contains four vertices connected by
three framed edges with K\"ahler parameters $t_1$, $t$ and
$t_2$~(Fig.~\ref{toricchiral}).
 Recall from Section~\ref{toric}
that the $\vec v_2$ and $\vec v_3^{\,\prime}$ edges of this toric
graph represent one-cycles in the fibers of the toric geometry of
$X_p$ and have length $t$. Gluing the open topological string
vertex to these edges thus corresponds to the insertion of
D-branes in the fiber of (\ref{Xfib}). In this local $\complex^3$
patch these are special Lagrangian submanifolds with the topology
of $\complex\times S^1$ corresponding to D-branes wrapping an
$S^1$ cycle in the fiber. The gluing edges labelled by $t_a$,
$a=1,2$ thus correspond to rational holomorphic curves
$\Sigma_a\cong\PP^1$ at distances $t_a$ from the sphere $\PP^1$
corresponding to the gluing edge labelled $t$ which represents the
base of the fibration (\ref{Xfib}). The $t_a$ themselves are the
K\"ahler parameters of the corresponding curves $\Sigma_a$.

This defines a nonsingular toric Calabi-Yau scheme $\hat X_p$. It
can be naturally thought of as emerging from a large $N$ geometric
transition from the toric threefold described by a nonplanar
trivalent graph with four vertices
\cite{Aganagic:2002qg}--\cite{Diaconescu:2002sf}. This geometry
locally contains two three-sphere cotangent bundles $T^*S^3$ each
constructed as a $\torus^2$ fibration over an interval and glued
together along a $\PP^1$, and it corresponds to a surgery
construction on $S^3$ wherein one performs a Heegaard split of
$S^3$ along solid tori and glues the tori together along their
boundary $\torus^2$ through the $SL(2,\zed)$ transformation that
relates one of the collapsing cycles of the toric geometry to the
other. Since this geometry is not globally the cotangent bundle of
a three-manifold, the topological string dynamics is described by
two $U(N)$ Chern-Simons theories on $S^3$, along with an
additional sector of open strings stretched between the two
three-spheres which correspond to non-degenerate holomorphic
instantons. The geometric transition consists of shrinking the two
three-spheres to points, and then resolving these singularities
with two copies of $\PP^1$ to realize the toric manifold $\hat
X_p$ depicted in Fig.~\ref{toricchiral}. This description exhibits
the two stacks of D-branes present in the geometry explicitly
(seen here as wrapping the Lagrangian submanifolds $S^3\subset
T^*S^3$ before the transition), and it may also account for the
generic discrepancy between the Chern-Simons and $q$-deformed
Yang-Mills descriptions of topological strings on $X_p$
\cite{Caporaso:2005ta}.

It is instructive to compare this result with the black hole partition
function in the sector of vanishing Ramond-Ramond flux. From the large
$N$ expansion of the $q$-deformed partition function for the full coupled
theory one finds from (\ref{AgZexpl}), (\ref{Zpmrel}) and
  (\ref{chiraltopvertex}) the result~\cite{Aganagic:2004js}
\bea Z^0_{\rm BH}&=& \big|{\sf Z}_0(q)\big|^2\,\sum_{\hat R_\pm}
\e^{-t |\hat R_+| -\bar{t}\,|\hat R_-| }\,(-1)^{p|\hat R_+|
+p|\hat R_-|}\,q^{(p-1)(\kappa_{\hat R_+} +\kappa_{\hat R_-})/2 }
\nonumber\\ && \times\,\sum_{\hat R_{(1)},\hat R_{(2)}}
\e^{-2\,{\rm Re}(t_1)|\hat R_{(1)}|}~ \e^{-{\rm Re}(t_2) |\hat
R_{(2)}|}\, (-1)^{|\hat R_{(1)}|+|\hat R_{(2)}|}\nonumber\\ &&
\times\, C_{0\hat R_{(1)}\hat R_+^\top}(q)\,  C_{0 \hat R_+\hat
R_{(2)}}(q)\, C_{0\hat R_-\hat R_{(1)}^\top}(q)\, C_{0\hat
R_-^\top\hat R_{(2)}^\top}(q) \ . \label{ZBH0full}\eea We see that
the chiral partition function (\ref{top3}) corresponds to the
contribution to (\ref{ZBH0full}) from the trivial sector $\hat
R_-=0$. This is natural in the toric description, as it
corresponds to dropping a gluing edge in the construction of the
corresponding threefold. Note that our K\"ahler modulus $t$ is
real as we have not included a $\theta$-angle in the definition of
the original $q$-deformed gauge theory.

\subsection{Gromov-Witten invariants\label{GWinvs}}

We will now analyse the Gromov-Witten invariants of the threefold
$\hat X_p$. Modulo torsion this space has, by construction, second
homology group $H_2(\hat X_p,\zed)\cong\zed^3$ with generating
two-cycles carrying the K\"ahler parameters $\mbf t:=(t_1,t,t_2)$.
Holomorphic string maps, which generate the chiral theory,
preserve orientations of curves and are classified by {\it
  effective} degrees taking values in $H_2(\hat
X_p,\zed)_+\cong\nat_0^3$. The
structure of eq.~(\ref{top3}) suggests that the chiral expansion of
the free energy ${\cal F}_{\rm chiral}^{q{\rm YM}}$ should be
  understood as a particular contribution to the more general free
  energy
 \beq
\label{gwgeneral}  {\cal F}_{\hat
X_p}=\sum_{g=0}^\infty\,g_s^{2g-2}~ \hat{\cal F}_g \eeq for
topological A-model strings on $\hat X_p$. The free energy at
genus $g$ is given by a sum over effective two-homology classes of
genus $g$ worldsheet instantons as \beq\label{gw1} \hat{\cal
F}_g(\mbf t;p)= \sum_{\mbf n\in H_2(\hat X_p,\zed)_+}\e^{-\mbf
n\cdot \mbf t}~ \hat{\rm N}^g{}^{~}_{\mbf n}(p) \ , \eeq where
$\hat{\rm N}^g{}^{~}_{\mbf n}(p)\in\rat$ are the Gromov-Witten
invariants of $\hat X_p$. The chiral amplitude is obtained by
taking $t_1=t_2=2\,\hat t$ and restricting the sums (\ref{gw1}) to
the sector where the effective degree $\mbf n:=(k_1,n,k_2)$ is
constrained by $k_1+k_2=k$ with $0\leq k\leq2n$. The
expansion of the free energy (\ref{gw1}) in the three independent
K\"ahler parameters can be used to easily
derive $\hat{\rm N}^g{}^{~}_{\mbf
  n}(p)$ and the generalized Gromov-Witten invariants ${\rm
  N}^g{}^{~}_{n,k}(p)$ of $X_p$ appearing in the large $N$ limit of chiral
$q$-deformed Yang-Mills theory through
\beq
\sum_{k_1=0}^k\,
\hat{\rm N}^g{}^{~}_{(k_1,n,k-k_1)}(p)={\rm N}^g{}^{~}_{n,k}(p) \ .
\label{hatNNrel}\eeq
In particular, $\hat{\rm
  N}^g{}^{~}_{(0,n,0)}(p)={\rm N}^g{}^{~}_{n}(p)$ are the
Gromov-Witten invariants of the original threefold $X_p$. We will
exhibit the first few invariants only for the
genus $g=0$ case. Higher genera are similarly dealt with.

The $n=1$ terms in eq.~(\ref{freen}) are of the form
$\sum_{k=0}^{2}\e^{-  t - 2\,k\,\hat t}~{\rm N}^g{}^{~}_{1,k}$.
The required two-homology classes in (\ref{gw1}) are given by \beq
\mbf n~=~(0,1,0) \ , \quad (1,1,0) \ , \quad (0,1,1) \ , \quad
(1,1,1) \ , \label{twohomvects}\eeq and the corresponding
Gromov-Witten invariants are \beq \hat{\rm N}^0{}_{(0,1,0)}=
-\hat{\rm N}^0{}_{(1,1,0)}=- \hat{\rm N}^0{}_{(0,1,1)}=\hat{\rm
N}^0{}_{(1,1,1)}=( -1)^{p} \ .\eeq For $n=2$, by expanding the sum
explicitly it is again easy to write down the contributing degrees
$\mbf n\in H_2(\hat X_p,\zed)_+$ and thus find the Gromov-Witten
invariants
 \bea
\hat{\rm N}^0{}_{(0, 2,0)}&=&\mbox{$\frac{1}{8}$}\,\left(1-4\,p+2\,p^2
\right) \ , \nonumber \\ \hat{\rm N}^0{}_{(1,2,0)}=
\hat{\rm N}^0{}_{(0,2,1)}&=& \mbox{$\frac{1}{2}$}\,\left(p-p^2
\right) \ , \nonumber\\ \nonumber
  \hat{\rm N}^0{}_{(0,2,2)}= \hat{\rm N}^0{}_{(2,2,0)}&=&-\mbox{$\frac{1}{8}$}\,\left(1-2\,p^2
\right) \ , \\ \nonumber
\hat{\rm N}^0{}_{(1,2,1)}&=&   p^2 \ ,\\ \nonumber
   \hat{\rm N}^0{}_{(2,2,1)}=\hat{\rm N}^0{}_{(1,2,2)}&=&
-\mbox{$\frac{1}{2}$}\,\left(p+p^2\right) \ , \\
\hat{\rm N}^0{}_{(2,2,2)}&=&\mbox{$\frac{1}{8}$}\,\left(1+4\,p+2\,p^2
\right) \ . \eea
Finally, by following the same route for $n=3$ we find the invariants
\bea \nonumber
\hat{\rm N}^0{}_{(0, 3,0)}&=&(-1)^p \,\left(\mbox{$\frac{1}{27}$}-
\mbox{$\frac{1}{3}$}\,p+\mbox{$\frac{5}{6}$}\,p^2-\mbox{$\frac{2}{3}$}\,
p^3+\mbox{$\frac{1}{6}$}\,p^4\right) \ , \\\nonumber
\hat{\rm N}^0{}_{(1,3,0)}=\hat{\rm N}^0{}_{(0,3,1)}&=&(-1)^p\,
\left(\mbox{$\frac{1}{6}$}\,p-
p^2+\mbox{$\frac{4}{3}$}\,p^3-\mbox{$\frac{1}{2}$}\,p^4\right) \ ,
\\\nonumber
 \hat{\rm N}^0{}_{(2,3,0)}=\hat{\rm N}^0{}_{(0,3,2)}&=& (-1)^p\,\left(
\mbox{$\frac{1}{6}$}\,p-\mbox{$\frac{2}{3}$}\,p^3+
\mbox{$\frac{1}{2}$}\,p^4\right) \ , \\\nonumber
\hat{\rm N}^0{}_{(1,3,1)}&=& (-1)^p\,
\left(\mbox{$\frac{1}{2}$}\,p^2-2\,p^3+\mbox{$\frac{3}{2}$}\,p^4
\right) \ , \\\nonumber \hat{\rm N}^0{}_{(3,
3,0)}=\hat{\rm N}^0{}_{(0,3,3)}&=&-(-1)^p\,
\left(\mbox{$\frac{1}{27}$}-\mbox{$\frac{1}{6}$}\,p^2+
\mbox{$\frac{1}{6}$}\,p^4\right)
\ , \\\nonumber\hat{\rm N}^0{}_{(1, 3,2)}&=&(-1)^p\,
\left(\mbox{$\frac{1}{2}$}\,p^2-\mbox{$\frac{3}{2}$}\,p^4
\right) \ , \\ \nonumber \hat{\rm N}^0{}_{(3,
3,1)}=\hat{\rm N}^0{}_{(1,3,3)}&=&-(-1)^p\,
\left(\mbox{$\frac{1}{6}$}\,p-\mbox{$\frac{2}{3}$}\,p^3-
\mbox{$\frac{1}{2}$}\,p^4\right) \ , \\\nonumber\hat{\rm N}^0{}_{(2,
3,2)}&=&(-1)^p\,\left(\mbox{$\frac{1}{2}$}\,p^2+2\,p^3+
\mbox{$\frac{3}{2}$}\,p^4\right) \ , \\ \nonumber
\hat{\rm N}^0{}_{(3, 3,2)}=\hat{\rm N}^0{}_{(2,3,3)}&=&-(-1)^p\,
\left(\mbox{$\frac{1}{6}$}\,p+p^2+\mbox{$\frac{4}{3}$}\,p^3+
\mbox{$\frac{1}{2}$}\,p^4\right) \ , \\
 \hat{\rm N}^0{}_{(3,3,3)}&=&(-1)^p\,\left(\mbox{$\frac{1}{27}$}+
\mbox{$\frac{1}{3}$}\,p+\mbox{$\frac{5}{6}$}\,p^2+
\mbox{$\frac{2}{3}$}\,p^3+\mbox{$\frac{1}{6}$}\,p^4\right) \ .
\eea

Let us now explore the hidden integrality structure of the
Gromov-Witten invariants through the embedding of the topological
A-model string theory into Type~IIA string theory on $\hat
X_p$~\cite{Gopakumar:1998ii,Gopakumar:1998jq}. The generating
function (\ref{gwgeneral}) for the all-genus topological string
amplitudes can be written as a generalized index that counts BPS
states of D2-branes wrapping holomorphic curves in $\hat X_p$ as
\bea \label{gv} {\cal F}_{\hat X_p}= \sum_{g=0}^{\infty}~
\sum_{\mbf n\in H_2(\hat X_p,\zed)_+}\, \hat{\rm n}^g{}^{~}_{\mbf
n}(p)\, \sum_{d=1}^\infty \,\frac{1}{d}\,\left( 2\,
\sin\big(\mbox{$\frac{d\,g_s}{2}$}\big)\right)^{2g-2}~\e^{-d\,\mbf
n\cdot \mbf t} \ , \eea where $\hat{\rm n}^g{}^{~}_{\mbf
n}(p)\in\zed$ are the Gopakumar-Vafa integer invariants which
compute the Euler characteristic of the moduli space of embedded
curves of genus $g$ and two-homology class $\mbf n$ in $\hat X_p$.
As a consistency test of our interpretation of the chiral gauge
theory as a topological string theory, we will now extract the
invariants $\hat{\rm n}^0{}^{~}_{\mbf n}(p)$ directly from the
expansion of the genus zero free energy and verify that they are
indeed integers. The case of higher genera can be similarly
handled by using the explicit inversion formula between the two
expansions (\ref{gwgeneral},\ref{gw1}) and (\ref{gv}) that
expresses the Gopakumar-Vafa invariants $\hat{\rm n}^g{}^{~}_{\mbf
n}$ in terms of Gromov-Witten invariants $\hat{\rm
N}^h{}^{~}_{\mbf m}$.

At genus zero we find by comparing (\ref{gwgeneral},\ref{gw1}) and
(\ref{gv}) the relation
 \bea \label{GWGV0rel}
\hat{\rm N}^0{}^{~}_{\mbf n}(p)= \sum_{d\,|\,\mbf n}\,\frac{1}{d^3}~
\hat{\rm n}^0{}^{~}_{\mbf n/d}(p) \ . \eea We can invert this expression for
a given effective class $\mbf n\in H_2(\hat X_p,\zed)_+$ to get the invariants
$\hat{\rm n}^0{}^{~}_{\mbf n}$. The first few identifications are simple.
For example, the degree $\mbf n=(1,1,0)$ has no divisors other than
$d=1$, and thus $\hat{\rm n}^0{}^{~}_{(1,1,0)}(p)=\hat{\rm
  N}^0{}^{~}_{(1,1,0)}(p)=-(-1)^p$. Instead, for the degree $\mbf
n=(2,2,0)$ we have
 \bea
\hat{\rm N}^0{}^{~}_{(2,2,0)}(p)=
\hat{\rm n}^0{}^{~}_{(2,2,0)}(p)+
\mbox{$\frac{1}{2^3}$}~\hat
{\rm n}^0{}^{~}_{(1,1,0)}(p)\eea
leading to
 \beq
\nonumber  \hat{\rm n}^0{}^{~}_{(2,2,0)}(p)=
-\mbox{$\frac{1}{8}$}\,\left(1-2\,p^2\right) +\mbox{$
\frac{1}{2^3}$}\, (-1)^{p}\eeq which is indeed an integer.
Proceeding iteratively along these same lines, it is
straightforward to derive the first few sets of Gopakumar-Vafa
invariants. Omitting those that are obtained by the obvious
symmetries, we obtain the list \bea\label{list1} \hat{\rm
n}^0{}^{~}_{(0,1,0)}&=& ( -1)^{p}=-\hat {\rm n}^0{}^{~}_{(1,1,0)}=
\hat{\rm n}^0{}^{~}_{(1,1,1)} \ , \\[10pt]
\hat{\rm n}^0{}^{~}_{(0,2,0)}&=&\mbox{$\frac{1}{8}$}\,\left(1-(-1)^p-4
\,p+2\,p^2\right) \ , \nonumber\\ \nonumber
\hat{\rm n}^0{}^{~}_{(0,2,1)}&=&\mbox{$\frac{1}{2}$}\,
\left(p-p^2\right) \ , \\ \nonumber
   \hat{\rm n}^0{}^{~}_{(2,2,0)}&=&-\mbox{$\frac{1}{8}$}\,\left(1-(-1)^p-2\,
   p^2\right) \ , \\\nonumber
   \hat{\rm n}^0{}^{~}_{(1,2,1)}&=&   p^2 \ , \\ \nonumber
   \hat{\rm n}^0{}^{~}_{(2,2,2)}&=&\mbox{$\frac{1}{8}$}\,
\left(1-(-1)^p+4\,p+2\, p^2\right) \ , \\
\hat{\rm n}^0{}^{~}_{(2,2,1)}&=&-\mbox{$\frac{1}{2}$}\,\left(p+p^2
\right) \ , \\[10pt]
\hat{\rm n}^0{}^{~}_{(0, 3,0)}&=&-(-1)^p\,\left(
\mbox{$\frac{1}{3}$}\,p-\mbox{$\frac{5}{6}$}\,p^2+
\mbox{$\frac{2}{3}$}\,p^3-\mbox{$\frac{1}{6}$}\,p^4
\right) \ , \nonumber\\
\hat{\rm n}^0{}^{~}_{(1,3,0)}&=&(-1)^p\,\left(\mbox{$\frac{1}{6}$}\,p-
p^2+\mbox{$\frac{4}{3}$}\,p^3-\mbox{$\frac{1}{2}$}\,p^4\right) \ ,
\nonumber\\ \hat{\rm n}^0{}^{~}_{(1,3,1)}&=& (-1)^p\,
\left(\mbox{$\frac{1}{2}$}\,p^2-2\,p^3+\mbox{$\frac{3}{2}$}\,p^4
\right) \ , \nonumber\\
\hat{\rm n}^0{}^{~}_{(2,3,0)}&=& (-1)^p\,
\left(\mbox{$\frac{1}{6}$}\,p-\mbox{$\frac{2}{3}$}\,p^3+
\mbox{$\frac{1}{2}$}\,p^4\right) \ , \nonumber\\
\hat{\rm n}^0{}^{~}_{(1, 3,2)}&=&(-1)^p\,
\left(\mbox{$\frac{1}{2}$}\,p^2-\mbox{$\frac{3}{2}$}\,p^4
\right) \ , \nonumber\\ \hat{\rm n}^0{}^{~}_{(3,
3,0)}&=&(-1)^p\,\left(\mbox{$\frac{1}{6}$}\,p^2-\mbox{$\frac{1}{6}$}\,
p^4\right) \ , \nonumber\\ \nonumber
\hat{\rm n}^0{}^{~}_{(2, 3,2)}&=&(-1)^p\,\left(\mbox{$\frac{1}{2}$}\,
p^2+2\,p^3+\mbox{$\frac{3}{2}$}\,p^4\right) \ , \\
\hat{\rm n}^0{}^{~}_{(3, 3,1)}&=&-(-1)^p\,\left(\mbox{$\frac{1}{6}$}\,
p-\mbox{$\frac{2}{3}$}\,p^3-\mbox{$\frac{1}{2}$}\,p^4\right) \ , \nonumber
\\ \nonumber \hat{\rm n}^0{}^{~}_{(3, 3,2)}&=&-(-1)^p\,
\left(\mbox{$\frac{1}{6}$}\,p+p^2+\mbox{$\frac{4}{3}$}\,p^3-
\mbox{$\frac{1}{2}$}\,p^4\right) \ , \nonumber\\
\hat{\rm n}^0{}^{~}_{(3,3,3)}&=&(-1)^p\,\left(
\mbox{$\frac{1}{3}$}\,p+\mbox{$\frac{5}{6}$}\,p^2+
\mbox{$\frac{2}{3}$}\,p^3+\mbox{$\frac{1}{6}$}\,p^4\right) \ .
\label{gwilist}\eea

Despite their appearance, one can easily check the integrality of
the expressions (\ref{list1})--(\ref{gwilist}) for any integer
$p$. The quantity $\hat{\rm n}^0{}^{~}_{(0,n,0)}(p)={\rm
n}^0{}^{~}_n(p)$ computes the genus zero Gopakumar-Vafa invariants
for closed topological strings on $X_p$. In particular, for $p=1$
the threefold $X_1=K^{1/2}_{\PP^1}\oplus K^{1/2}_{\PP^1}$ is the
resolved conifold (with $K_{\PP^1}$ the canonical line bundle over
$\PP^1$) which may be described using the large $N$ geometric
transition from $U(N)$ Chern-Simons theory on $S^3$
\cite{Gopakumar:1998ki}. In this case the Gopakumar-Vafa
invariants are simply given by ${\rm
n}^0{}^{~}_n(1)=-\delta_{n,1}$, which is the well-known result.
Similarly, for $p=2$ (and also $p=0$) one has $X_2={\cal O}\oplus
K_{\PP^1}$ (with $\cal O$ the trivial
  line bundle over $\PP^1$) and one finds the
invariants ${\rm n}^0{}^{~}_n(2)=\delta_{n,1}$, again as
expected~\cite{bryankatzleung}. These patterns can be further
verified by extending the computations above up to seventh order
(see Appendix~\ref{GopVafainvs}). For $p\geq3$ our calculations
suggest that there are infinitely many non-vanishing
Gopakumar-Vafa invariants of $X_p$. As we will see in
Section~\ref{stringseries}, this observation is related to the
finite radius of convergence of the topological string
perturbation series for $p>2$ and it suggests an ultimate reason
for the different phase structures of the theories with $p=0,1,2$
and $p>2$.

\subsection{Gross-Taylor expansion as a topological string
  theory\label{GTexp}}

We will now establish the connection between topological string theory
and the Gross-Taylor string expansion. By specializing to the genus
zero contributions, we will further establish the connection with the
Crescimanno-Taylor \cite{Crescimanno:1994eg} expansion which will be used extensively in the
next section. Recall from Section~\ref{Chiralexp} that ordinary
Yang-Mills theory on $S^2$ is reached in the limit where
$p\to\infty$ with the area combination $g_sN\,p=A$ fixed. To describe
this limit, we first rewrite the chiral partition function
(\ref{ZqYMchiral0}) in terms of the relevant parameters as
\bea
{\cal Z}_{\rm chiral}^{q{\rm YM},0}&=&
\sum_{n=1}^{\infty}\e^{-(\frac{1}{2}-\frac{1}{p})n\,A}\,\sum_{\hat
R}\e^{-{\kappa_{\hat R}A}/{2\,N}}\,W_{\hat
R}\big(\e^{{A}/{p\,N}}\big)^2\nonumber\\ && \qquad\qquad\qquad\qquad\quad
\times\,\prod^{i_{\rm max}}_{i=1}~
\prod_{j=1}^{n_i}\left(1-\e^{-\frac{A}{p\,N}\,(i-j)}~
\e^{-{A}/{p}}\right)^2 \ . \label{Zchiralrel}
\eea

The crucial observation is that the quantum dimension
(\ref{quantdimconv}) tends smoothly to the ordinary classical dimension
$\dim(\hat R)$ of the $SU(N)$ representation $\hat R$ in the limit
$g_s\to 0$. This means that, in the double limit $N\to\infty,\,p\to\infty$, the
factor $\dim_q(\hat R)^2$ must contain terms
reproducing the expansion of the classical dimension
\beq
\dim\big(\hat R\big)=\xi_{2n}\,N^{2n}+\xi_{2n-1}\,N^{2n-1}+ \dots \ .
\eeq
These terms correspond to the contributions
\beq
\dim_q\big(\hat R\big)=\xi_{2n}\,\left(\mbox{$\frac{p\,N}{A}$}
\right)^{2n}\,\left(\mbox{$\frac{A}{p}$}\right)^{2n}+
\xi_{2n-1}\,\left(\mbox{$\frac{p\,N}{A}$}\right)^{2n-1}\,
\left(\mbox{$\frac{A}{p}$}\right)^{2n-1}+\dots \ .
\label{qdimcontr}\eeq
The subleading terms as $g_s\to0$ are of the form
$g_s^m\,(g_sN)^k=(\frac{A}{p\,N})^{m}\,(\frac{A}{p})^{k}$
with $m+k>0$ by smoothness of the classical
limit. From the explicit expression in (\ref{Zchiralrel}) it follows
that $k\geq 0$. If $m<0$ the corrections are
$(\frac{p\,N}{A})^{|m|}\,(\frac{A}{p})^{k}=N^{|m|}\,(\frac{A}{p})^{k-|m|}$
and therefore correspond to $\frac Ap$ corrections in
(\ref{qdimcontr}) to the classical terms given by
\beq
\dim_q\big(\hat R\big)=\xi_{2n}\,N^{2n}\,\left(1+\alpha_{2n}\,
\mbox{$\frac
  Ap$}+\dots\right)+\xi_{2n-1}\,N^{2n-1}\,\left(1+\alpha_{2n-1}\,
\mbox{$\frac Ap$}+\dots\right)+\dots \ .
\label{Apcorrsdimq}\eeq
On the other hand, when $m>0$ there are corrections of the
type $\frac{1}{N^m}\,(\frac{A}{p})^{k+m}$ which correspond to both
``quantum'' contributions to the genus expansion and to $\frac Ap$
corrections. These terms do not appear at leading order $N^2$ in the
free energy.

We can now explicitly match the $p\to\infty$ limit of the chiral
topological string expansion with the usual chiral Gross-Taylor \cite{Gross:1993hu} series
at leading order. The genus zero chiral free energy of the topological
string is given by
\beq {\cal F}_0\big(t\,,\,\hat t\,;\,p\big)
=\frac{1}{g_s^2}\,
\sum_{n=1}^{\infty}\e^{-n\,t}\,
\sum_{k=0}^{2n}\e^{-2\,k\,\hat t}\,\sum_{k_1=0}^k\,
\hat{\rm N}^{0}{}^{~}_{(k_1,n,k-k_1)}(p)\label{gsincl}\eeq
in terms of Gromov-Witten invariants of $\hat X_p$.
By rewriting this expansion in terms of QCD$_2$
parameters and expanding in $\frac1p$ we get \beq {\cal
  F}_0(A;p)=N^2\,
\sum_{n=1}^{\infty}\e^{-{n\,A}/{2}}\,
\sum_{k=0}^{2n}~\sum_{l,m=1}^{\infty}\,
\frac{k^l\,n^m}{l!\,m!}\,\left(-\frac{A}{p}\right)^{l+m-2}\,
\sum_{k_1=0}^k\,\hat{\rm N}^{0}{}^{~}_{(k_1,n,k-k_1)}(p)
\ . \label{top1p}\eeq On the other hand, the
genus zero chiral free energy of the QCD$_2$ string can be written in
the form \beq
\Phi_0(A) =N^2\,
\sum_{n=1}^{\infty}\e^{-{n\,A}/{2}}\,
\sum_{j=0}^{2n-2}\eta^{(n)}_j~A^j \label{GW0}\eeq
since the branch point and $\Omega$-point singularities of the
holomorphic string maps generate,
at winding number $n$, a polynomial of degree $2n-2$ in the area $A$. The
matching of (\ref{top1p}) and (\ref{GW0}) thereby determines the polynomial
coefficients $\eta^{(n)}_j$ in terms of the leading behaviour as
$p\to\infty$ of Gromov-Witten invariants through
\beq
\eta^{(n)}_j=\frac1{(j+2)!}~\lim_{p\to\infty}\,(-p)^{-j}\,
\sum_{k=0}^{2n}\,(n+k)^{j+2}\,\sum_{k_1=0}^k\,
\hat{\rm N}^{0}{}^{~}_{(k_1,n,k-k_1)}(p)
 \ . \label{HurGWrel}\eeq

This relationship suggests an explicit realization of two-dimensional
chiral Yang-Mills theory as a topological string theory. From a
physical perspective, as the degree of ${\cal O}(-p)\to\PP^1$ grows
the higher Kaluza-Klein modes of any section of this line
bundle decouple and we may formally identify ${\cal O}(-\infty)$ with
the trivial line bundle $\PP^1\times\complex$. Thus the Gromov-Witten
invariants of $X_\infty$ formally reproduce the counting of
holomorphic maps into the base sphere. We will see some explicit
examples of this in Section~\ref{stringseries}. Note that the D-brane
insertions in the fibers of the original fibration (\ref{Xfib}) play a
crucial role in this identification.

Let us now describe the mathematical implications of the relationship
(\ref{HurGWrel}). Let $\overline{M}_g(\hat X_p,\mbf n)$ be the
Deligne-Mumford moduli space of stable holomorphic maps from connected
genus $g$ curves to $\hat X_p$ which represent the class $\mbf n\in
H_2(\hat X_p,\zed)$. Then the Gromov-Witten invariants of $\hat X_p$
are given by
\beq
\hat{\rm N}^g{}^{~}_{\mbf n}(p)=\int_{\,\overline{M}_g(\hat X_p,\mbf
  n)}~1 \ .
\label{GWhatXpdef}\eeq
More precisely, the integral should be evaluated over the virtual
fundamental class of the moduli space of maps. With the
appropriate push-forward map one can use virtual
localization techniques to reduce the integral (\ref{GWhatXpdef}) to
Hodge integrals over the moduli space of  curves with $n$ punctures
$\overline{M}_{g,n}$ of dimension $3g-3+n$. While this is always
possible to do in principle, in practise it is quite difficult.

On the other hand, the relationship (\ref{HurGWrel}) gives such a
reduction in the limit $p\to\infty$ by relating (\ref{GWhatXpdef}) to
the Gromov-Witten theory of the base $\PP^1$. For this, let ${\rm
  H}^g{}^{~}_{\vec\mu}$ be the Hurwitz numbers of $\PP^1$
corresponding to the partition $\vec\mu=(1^{\mu_1}\,2^{\mu_2}\cdots)$,
i.e. the number of genus $g$ branched covering maps to $\PP^1$ with
ramification $\vec\mu$ over $\infty$ and simple ramification over
$\PP^1\,\setminus\,\infty$. These integers can be represented as integrals
over the moduli space $\overline{M}_g(\PP^1,n)$ of holomorphic maps of
genus $g$ and winding number $n$ to $\PP^1$. Let us illustrate how
this works explicitly in the simplest case of the trivial partition
$\vec\mu=(1^n)$, i.e. the case of genus $g$ simple branched covers of
degree $n$ over $\PP^1$. By the Riemann-Hurwitz theorem, such maps
have $r=2g-2+2n$ simple ramification points. There is a
natural map $\beta:\overline{M}_g(\PP^1,n)\to\PP^r$ which assigns to
each map its branch point locus. Let $\Xi$ be the hyperplane class of
$\PP^r$ associated to the canonical hyperplane bundle. Then the
simple Hurwitz numbers can be represented as Gromov-Witten integrals
 \cite{Okounkov:2000gx}--\cite{vainshtein}
\beq
{\rm H}^g{}^{~}_n:={\rm H}^g{}^{~}_{(1^n)}=
\int_{\,\overline{M}_g(\PP^1,n)}~\beta^*(\Xi) \ .
\label{simpleHurGW}\eeq

The virtual localization formula may now be used to compute the
integral (\ref{simpleHurGW}). The standard action of the multiplicative
group of complex numbers $\complex^\times$ on $\PP^1$ induces a
$\complex^\times$-action on the moduli space
$\overline{M}_g(\PP^1,n)$ for which the pullback $\beta^*(\Xi)$ is an
equivariant class. The fixed points of this group action are products
of moduli spaces of $n$-punctured curves $\overline{M}_{g,n}$. The
localization formula thereby reduces (\ref{simpleHurGW}) to
tautological intersection indices on $\overline{M}_{g,n}$. The result
can be expressed as a Hodge integral as follows. Let $\mathcal{L}_i$,
$i=1,\dots,n$ be the canonical line bundles over $\overline{M}_{g,n}$,
and define tautological classes as the first Chern classes
$\psi_i=c_1(\mathcal{L}_i)\in H^2(\,\overline{M}_{g,n},\rat)$. Let
$\mathcal{E}_g\to\overline{M}_{g,n}$ be the rank~$g$ Hodge bundle, and
denote the corresponding Chern classes by
$\lambda_k=c_k(\mathcal{E}_g)\in
H^{2k}(\,\overline{M}_{g,n},\rat)$. With $\lambda_0:=1$, the
localization formula then reads  \cite{Okounkov:2000gx}--\cite{vainshtein}
\beq
{\rm H}^g{}^{~}_n=\frac{(2g-2+2n)!}{n!}\,\sum_{k=0}^g\,(-1)^k\,
\int_{\,\overline{M}_{g,n}}\lambda_k\wedge\bigwedge_{i=1}^n\,
(1-\psi_i)^{-1}
\label{locformula}\eeq
for $(g,n)\neq(0,1),(0,2)$.

On the other hand, the coefficient of $A^{2n-2}/(2n-2)!$ in the
area polynomials of (\ref{GW0}) is precisely the number of
topologically inequivalent holomorphic maps $\PP^1\to\PP^1$ with
$2n-2$ branch point singularities of fixed image and no
$\Omega$-point singularities. Thus ${\rm
  H}^0{}^{~}_n=(2n-2)!\,\eta_{2n-2}^{(n)}$. Using (\ref{HurGWrel}) we
may thereby write a large $p$ localization formula for the
Gromov-Witten invariants (\ref{GWhatXpdef}) as
\beq
\lim_{p\to\infty}\,\frac1{p^{2n-2}}\,\sum_{k=0}^{2n}\,
(n+k)^{2n}\,\sum_{k_1=0}^k\,\int_{\,
\overline{M}_0\big(\,\hat X_p\,,\,(k_1,n,k-k_1)\big)}~1
=\frac{(2n)!}{n!}\,
\int_{\,\overline{M}_{0,n}}\,\bigwedge_{i=1}^n\,
(1-\psi_i)^{-1}
\label{GWhatXploc}\eeq
for $n\neq1,2$. The right-hand side of eq.~(\ref{GWhatXploc}) can be
evaluated explicitly in these simple instances in terms of the Hodge
integrals $\sum_{a_k}\,\int_{\,\overline{M}_{0,n}}\,\psi_1^{\wedge
  a_1}\wedge\cdots\wedge\psi_n^{\wedge a_n}$, giving the anticipated genus
zero Hurwitz formula ${\rm H}^0{}^{~}_{n}=(2n-2)!\,n^{n-3}/n!$.

With some work one can extend these identifications to both higher
genera ($g>0$) and to covering maps with non-simple ramifications
of their branch point singularities ($j<2n-2+2g$). In these cases
one must take into account the contributions from $\Omega$-point
singularities in the above analysis. In the chiral gauge theory,
the singularity at an $\Omega$-point is a multiple branch point
singularity which is described by an arbitrary permutation on the
sheets of the covering space when following the lift of a closed
target space curve on $\PP^1$ around the $\Omega$-point. These
additional singularities feature in nicely in what is known about
the evaluation of the corresponding localization integrals in
Gromov-Witten theory. Any continuous mapping from a Riemann
surface to $\PP^1$ is, up to homotopy, the composition of a pinch
map (collapsing regions of the surface to a single point) and a
branched covering map. The pinch maps are responsible for the
$\Omega$-point singularities and are related to appearance of {\it
multiple} Hurwitz numbers (ramified over more than one point than
just $\infty$ on $\PP^1$) in the computation of the Gromov-Witten
invariants of the original threefold $X_p$ \cite{Bryan:2004iq}.
Physically, they are directly related to the insertions of the
fiber D-branes in the topological string theory on $X_p$. We will
return to the relationship between Hurwitz numbers and
Gromov-Witten invariants in Section~\ref{stringseries}.

\section{Matrix model formalism\label{saddle}}

In this section we shall investigate the large $N$ limit of the
chiral sector of $q$-deformed Yang-Mills theory on the sphere $S^2$ by
means of matrix model techniques. From a detailed analysis of the
resulting saddle-point equation we will obtain the phase structure of
the gauge theory at large $N$. We then show that in an appropriate
strong-coupling phase the chiral topological string theory of the
previous section is recovered from the large $N$ gauge dynamics.

\subsection{Saddle point solution\label{saddlepteq}}

\begin{figure}[htb]
\label{cut1}
\begin{center}
\epsfxsize=4.0 in \epsfbox{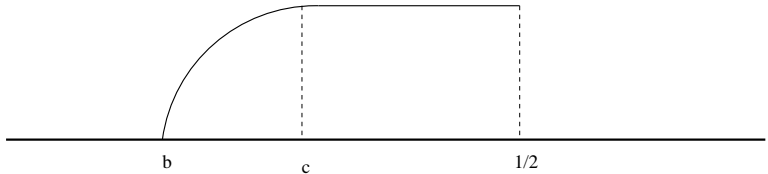}  \epsfysize=2.0 in
\end{center}
\caption{Ansatz for the distribution function $\rho(z)$.}
\label{onecutchiral}\end{figure}

The saddle point equation governing the distribution of Young tableaux
variables in chiral $q$-deformed Yang-Mills theory coincides with that
of the full coupled gauge theory
\cite{Caporaso:2005ta,Arsiwalla:2005jb,Jafferis:2005jd}. 
The new information is completely encoded in boundary conditions on
the solutions to this 
equation. In the chiral sector we sum not over all
representations but only over those
which have finitely many ``large'' numbers of rows. With $x_i:=\frac
iN$, this means that the lengths of the rows satisfy the constraints
\beq -\frac{n_i}N+x_i-\frac12=0 \ , \quad i\ge k\eeq
for some $k$ such that $x_k\to c_0\in\real$ as
$N\to\infty$. Equivalently, in the large $N$ limit we can characterize the
chiral theory by setting \beq n(x)=x-\mbox{$\frac{1}{2}$}  \ , \quad
x\ge c_0 \eeq where $n(x)$ is the number of boxes at position $x$
which is a monotonic function with $n(x_i)=\frac{n_i}N$. In
terms of the distribution function $\rho(x)$ for the Young tableaux,
this constraint can be written as \beq
\rho(n):=\frac{\dd x(n)}{\dd n}=1 \ , \quad
c=c_0-\mbox{$\frac12$}\le n\le\mbox{$\frac12$} \ . \eeq
In the matrix model approach we therefore have to solve the saddle-point
equation  \cite{Caporaso:2005ta,Arsiwalla:2005jb,Jafferis:2005jd}
\beq \label{cippino5bis} \frac{A\,
z}2=\hat t\,\int_b^{1/2} \dd w~\rho(w)\,\coth\bigl(\hat t\,(z-w)\bigr)
\ , \eeq
where $A=2\,\hat t\,p=N\,g_sp$ is the area parameter introduced
previously and the boundary conditions for $\rho(z)$ are depicted in
Fig.~\ref{onecutchiral}.
\medskip

Since $\rho(z)=1$ for $z\in[c,\frac12]$, eq.
(\ref{cippino5bis}) can be cast in the more standard form \beq
\label{piou}
\frac{A\,z}2=\hat t\,\int_b^{c} \dd w~\tilde\rho(w)\,\coth\bigl(\hat t\,(z-w)
\bigr)-\log\left|\mbox{$\frac{\sinh\bigl(\hat t\,(z-\frac12)
\bigr)}{\sinh\bigl(\hat t\,(z-c)\bigr)}$}\right|
\eeq where $\tilde \rho:=\rho|_{[b,c]}$. The new boundary conditions are now
translated into a $c$-dependent modification of the original
potential. This apparently mild modification is similar to the one
which occurs in the two-cut solution of the full non-chiral theory and it
will play an important role in recovering the correct perturbative
string expansion. To solve eq. (\ref{piou}), we first reduce it to
a classical Riemann-Hilbert problem by changing variables from $w$
and $z$ to $s=\e^{2\,\hat t\,w+8\,\hat t^{\,2} c/A}$ and
$u=\e^{2\,t\,z+8\,\hat t^{\,2} c/A}$ to get
\bea \frac{A}{8\,\hat t^{\,2}}\,\frac{\log(s)}{s}+\frac{2}{\hat t\,s}
\log\left|\mbox{$\frac{s
-\e^{\hat t+{8\,\hat t^{\,2} c}/A}}{s-
\e^{2\,\hat t \,c +{8\,\hat t^{\,2} c}/A}}$}\right|=
\int_{\e^{2\,\hat t\,b+8\,\hat t^{\,2} c /A}}^{\e^{2\,\hat t\,c
    +8\,\hat t^{\,2} c /A}}{\dd u}~\frac{
\varrho(u)}{{s}-{u}} \ , \label{RHproblem}\eea where \beq
\varrho(u):=\frac{\tilde\rho\bigl(\log(\e^{-8\,\hat t^{\,2} c/A}\, u
  )/2\,\hat t\,\bigr)}{2\,\hat t\,u} \ . \label{varrhodef}\eeq
Instead of the pair $(A,\hat t\,)$ we shall use the variables
$(A,p)$, as this will enable a simpler comparison with the
Gross-Taylor series later on. We also introduce the parameters
\beq b^\prime=\mbox{$\frac A{p^2}$}\,(p\,b+2\, c) \ ,\quad
c^\prime=\mbox{$\frac{A\,
c}{p^2}$}\,(p+2) \ , \quad d^\prime=\mbox{$\frac{2\,A
}{p^2}\,\left(c+\frac{p}{4}\right)$} \eeq in order to simplify
notation.

\begin{figure}[htb]
\begin{center}
\epsfxsize=4.0 in \epsfbox{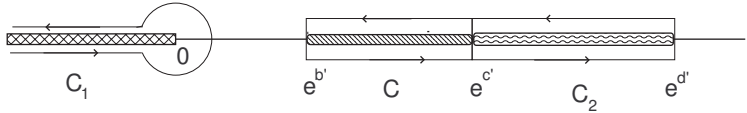}
\end{center}
\caption{The contour $C_1$ surrounds the branch cut $(-\infty,0]$ of
  $\log(z)$, $C_2$ encircles the cut $[\e^{c^\prime},\e^{d^\prime}]$
  of $\log\big(\frac{z-\e^{d^\prime}}{z-\e^{ c^\prime}}\big)$, while $C$
  encloses the physical branch cut $[\e^{b^\prime},\e^{c^\prime}]$ of
 $\sqrt{(z-\e^{b^\prime})(z-\e^{c^\prime})}$.}
\label{fg}\end{figure}

The solution to the integral
equation (\ref{RHproblem}) can then be written in terms of the
corresponding resolvent function as \bea \omega(z):=
\int_{\e^{b'}}^{\e^{c'}}\dd u~
\frac{\varrho(u)}{z-u} &=&\frac{p^2}
{4\pi\ii A}\,\sqrt{(z-\e^{c^\prime})(z-\e^{b^\prime})}\,
\nonumber\\ && \times\,\oint_{C}\, \frac{\dd w}{w(w-z)}~
\frac{{\log(w)}+\frac{2 }{p }\,\log\left|\frac{w -\e^{d^\prime}}{w-
 \e^{c^\prime}}\right|}{\sqrt{(w-\e^{b^\prime})(\e^{c^\prime}-w)}} \
, \label{resfndef}\eea where the closed contour $C$ encircles the support
$[\e^{b'},\e^{c'}]$ of the distribution $\varrho(u)$ with
counterclockwise orientation in the complex $z$-plane. If we choose
the square root and logarithmic branch cuts in (\ref{resfndef}) as
indicated in Fig.~\ref{fg},
then since the integrand decays as $w^{-3}$ at $|w|\to\infty$ we can deform
the contour of integration $C$ so that it encircles the cuts of the two
logarithms. This deformation picks up an additional contribution
from the pole at $w=z$ and we find \bea \omega(z)&=&-\frac{ p^2}
{4\pi\ii A}\,\sqrt{(z-\e^{c^\prime})(z-\e^{b^\prime})}\,\left[
\oint_{C_1}\,\frac{\dd w}{w(w-z)}~\frac{{\log(w)+\frac{2 }{p
}\,\log\left(\frac{w -\e^{d^\prime}}{w-
 \e^{c^\prime}}\right)}}{\sqrt{(w-\e^{c^\prime})(w-\e^{b^\prime})}}
\right.\nonumber\\ &&+\left.
 \oint_{C_2}\,\frac{\dd w}{w(w-z)}~\frac{\log(w)+\frac{2 }{p }\,
 \log\left(\frac{w-\e^{d^\prime}}{w-
 \e^{c^\prime}}\right)}{\sqrt{(w-\e^{c^\prime})(w-\e^{b^\prime})}}
\right]-\frac{ p^2}{2\,A}\,\frac{\log(z)}{z}-
 \frac{p }{A\,z}\,\log\left(\mbox{$\frac{z-\e^{d^\prime}}{z-
     \e^{c^\prime}}$}\right)\nonumber\\[5pt]
&=&-\frac{p^2}{2\,A}\,\sqrt{(z-\e^{c^\prime})(z-\e^{b^\prime})}\,
\left[\int^{-\epsilon}_{-\infty}\,\frac{\dd
  w}{w(w-z)}~\frac{1}{\sqrt{(\e^{c^\prime}-w)(\e^{b^\prime}-w)}}
\right.\nonumber\\ &&+\left.  \frac{\log(\epsilon)}{z}~
\e^{-(b^\prime+c^\prime\,)/2}-\frac{2}{p}\,
 \int_{\e^{c^\prime}}^{\e^{d^\prime}}\,\frac{\dd w}{w(w-z)}~
\frac{1}{\sqrt{(w-\e^{c^\prime})(w-\e^{b^\prime})}}\right]
\\ &&-\frac{ p^2}{2\,A}\,\frac{\log(z)}{z}-
 \frac{p }{A\,z}\,\log\left(\mbox{$\frac{z-\e^{d^\prime}}{z-
       \e^{c^\prime}}$}\right)-
\frac{p}{A\,z}\,\left(d^\prime-c^\prime\,\right)~
\e^{-(b^\prime+c^\prime\,)/2}\,\sqrt{(z-\e^{c^\prime})(z-\e^{b^\prime})}
\ , \nonumber
\eea where $\epsilon\to0^+$ and we have chosen positive square roots
for the integrands over the real line.

The required integrals are computed in Appendix~\ref{ChiralInts} and
one arrives at the resolvent
\bea
&& \omega(z)= -\frac{p}{2\,A\,
s}\,\biggl\{2\,p\,\log\left(\mbox{$\frac{\e^{{b^\prime}/{2}}\,
\sqrt{z-\e^{c^\prime}}
+\e^{{c^\prime}/{2}}\,\sqrt{z-\e^{b^\prime}}}{\sqrt{z-\e^{c^\prime}}
+\sqrt{z-\e^{b^\prime}}}$}~\right)\biggr.\nonumber\\ && \quad+\,p~
\e^{-{(b^\prime+c^\prime\,)}/{2}}\,
\sqrt{(z-\e^{c^\prime})(z-\e^{b^\prime})}\,
    \left( b^\prime+c^\prime- 2\,\log\bigl(\mbox{$\frac{\e^{{b^\prime}/{2}}
          +\e^{{c^\prime}/{2}}}2$}\bigr)\right)\nonumber\\ && \quad+\,2\,
\log\left(\mbox{$\frac{\left(\sqrt{\e^{d^\prime}-\e^{b^\prime}}+
\sqrt{\frac{z-\e^{b^\prime}}{z-\e^{c^\prime}}\,(\e^{d^\prime}-
\e^{c^\prime})}~\right)^2}{\e^{c^\prime}-\e^{b^\prime}}$}\right)\label{ckl}\\
&&\quad+\biggl.2~\e^{-{(b^\prime+c^\prime\,)}/{2}}\,
\sqrt{(z-\e^{c^\prime})(z-\e^{b^\prime})}\,\left[
 \log\left(\mbox{$\frac{\left(\e^{{c^\prime}/{2}}\,
\sqrt{\e^{d^\prime}-\e^{b^\prime}}+
\e^{{b^\prime}/{2}}\,\sqrt{\e^{d^\prime}-\e^{c^\prime}}~\right)^2}
{\e^{c^\prime}-\e^{b^\prime}}$}\right)-c^\prime\right]\biggr\} \ . \nonumber
\eea The asymptotic boundary condition for $\omega(z)$ is
fixed by the normalization of the spectral density to be \beq \label{ckl1}
\lim_{|z|\to\infty}\,\omega(z)=\frac{1}{z}\,
\int_{\e^{b^\prime}}^{\e^{c^\prime}} \dd u~
\varrho(u)=\frac{\mbox{$\frac{1}{2}$}+c}{z} \ , \eeq
which on comparison with the asymptotic
behaviour of (\ref{ckl}) yields a pair of equations for the unknown
parameters $b$ and $c$ of the saddle point solution. The constant
asymptotic value of (\ref{ckl}) at infinity is given by all terms multiplying
the square root $\sqrt{(z-\e^{c^\prime})(z-\e^{b^\prime})}$. Requiring
them to vanish imposes the boundary condition\beq \label{qTC1}
\frac{p}{2}\,\left(
b^\prime+c^\prime- 2\,\log\bigl(\mbox{$\frac{\e^{{b^\prime}/{2}} +
\e^{{c^\prime}/{2}}}{2}$}\bigr) \right)+
\log\left(\mbox{$\frac{\left(\e^{{c^\prime}/{2}}\,\sqrt{\e^{d^\prime}-\e^{b^\prime}}+
\e^{{b^\prime}/{2}}\,\sqrt{\e^{d^\prime}-\e^{c^\prime}}~\right)^2}
{\e^{c^\prime}-\e^{b^\prime}}$}\right)=c^\prime \ .
\eeq
Extracting the subleading behaviour of (\ref{ckl}) at $|z|\to\infty$
requires just a bit more effort and is fixed by eq.~(\ref{ckl1}) to be
 \beq
 \label{qTC2}
p\,\log\left(\mbox{$\frac{\e^{{b^\prime}/{2}}
+\e^{{c^\prime}/{2}}}{2}$}\right)+
\log\left(\mbox{$\frac{\left(\sqrt{\e^{d^\prime}-\e^{b^\prime}}+
\sqrt{\e^{d^\prime}-\e^{c^\prime}}~\right)^2}
{\e^{c^\prime}-\e^{b^\prime}}$}\right)=-\frac{A}{p}\,\left(
\mbox{$\frac{1}{2}$}+c\right) \ ,
\eeq where we have dropped a term which vanishes by eq.~(\ref{qTC1}).

When the equations above for the endpoints of the support interval hold,
all the terms in $\omega(z)$ which are proportional to
$\sqrt{(z-\e^{c^\prime})(z-\e^{b^\prime})}$ vanish identically, and
thus the resolvent function (\ref{ckl}) assumes the very simple and
compact form \bea
\omega(z)&=&-\frac{p^2}{A\,
z}\,\log\left(\mbox{$\frac{\e^{{b^\prime}/{2}}\,\sqrt{z-\e^{c^\prime}}
+\e^{{c^\prime}/{2}}\,\sqrt{z-\e^{b^\prime}}}{\sqrt{z-\e^{c^\prime}}
+\sqrt{z-\e^{b^\prime}}}$}~\right)\nonumber\\ && -\,\frac{p}{A\,z}\,
\log\left(\mbox{$\frac{\left(\sqrt{\e^{d^\prime}-\e^{b^\prime}}+
\sqrt{\frac{z-\e^{b^\prime}}{z-\e^{c^\prime}}\,(\e^{d^\prime}-
\e^{c^\prime})}~\right)^2}{\e^{c^\prime}-\e^{b^\prime}}$}\right) \ .
\label{cklcomp}\eea The distribution function (\ref{varrhodef}) is
then determined by the jump discontinuity of (\ref{cklcomp}) across
the cut $[\e^{b^\prime},\e^{c^\prime}]$ as
\bea\varrho(s)&=&\frac{\omega(s+\ii\epsilon)-\omega(s-\ii\epsilon)}{2\pi
\ii}\nonumber\\[5pt] &=&\frac{p^2}{\pi\,A\,
s}\,\arctan\left(\mbox{$\sqrt{\frac{\e^{c^\prime}-s}{s-\e^{b^\prime}}}$}~
\right)-\frac{p^2}{\pi\,A\,
s}\,\arctan\left(\mbox{$\e^{{b^\prime}/{2}-
{c^\prime}/{2}}\,\sqrt{\frac{\e^{c^\prime}-s}{s-\e^{b^\prime}}}$}~
\right)\nonumber\\ &&+\,\frac{2\,
p}{\pi\,A\,s}\,
\arctan\left(\mbox{$\sqrt{\frac{s-\e^{b^\prime}}{\e^{c^\prime}-s}\,
\frac{\e^{d^\prime}-\e^{c^\prime}}{\e^{d^\prime}-\e^{b^\prime}}}$}~\right)
\ . \label{1cutexpl}
\eea This distribution manifestly satisfies the required boundary
conditions, since $\varrho(\e^{c^\prime})=p/A\,s$ and
$\varrho(\e^{b^\prime})=0$.

\subsection{Crescimanno-Taylor equations\label{Cteqs}}

Eqs.~(\ref{qTC1}) and (\ref{qTC2}) encode most of the gauge dynamics
of the large $N$ limit. Before attempting a systematic solution, we
shall verify that they are consistent with the Crescimanno-Taylor
equations for the QCD$_2$ string \cite{Crescimanno:1994eg}, which will also provide a non-trivial
check of our equations. The usual chiral gauge theory on $S^2$ should
emerge in our framework when $p\to \infty$ with the area $A$ of the
sphere fixed. By substituting in the large $p$ expansions \bea
\sqrt{\e^{d^\prime}-\e^{c^\prime}}&=&
\left(\mbox{$\frac{A}{p}$}\right)^{1/2}\,\sqrt{\mbox{$\frac{1}{2}$}
-c}+\dots \ , \nonumber\\
\sqrt{\e^{d^\prime}-\e^{b^\prime}}&=&
\left(\mbox{$\frac{A}{p}$}\right)^{1/2}\,\sqrt{\mbox{$\frac{1}{2}$}
-b}+\dots \ , \nonumber\\
\e^{c^\prime}-\e^{b^\prime}&=&
\mbox{$\frac{A}{p}$}\,(c-b)+\dots \ , \nonumber\\
\log\left(\mbox{$\frac{\e^{{b^\prime}/{2}}+\e^{{c^\prime}/{2}}}{2}$}
\right)&=&\mbox{$\frac{A}{2\,p}$}\,(b+c)+\dots \eea
the expression (\ref{qTC1}) takes the form
 \bea \label{yty} &&-\frac{A}{4}\,(b+c)=\log\left(\mbox{$
\frac{\left( \sqrt{\frac{1}{2}-c}+
 \sqrt{\frac{1}{2}-b}~\right)^2}{(c-b)}$}\right) \ .
\eea This is the first Crescimanno-Taylor equation. Performing the
same expansion in eq.~(\ref{qTC2}) at leading order again recovers
eq.~(\ref{yty}).

The second Crescimanno-Taylor equation appears by
expanding either of eqs.~(\ref{qTC1}) or~(\ref{qTC2}) to second order
in $\frac1p$, giving the same result \beq
\mbox{$\frac{A}{16}$}\,{\left( b - c \right) }^2   + \,
{\sqrt{\left(\mbox{$\frac{1}{2}$} - b \right)\left(
\mbox{$\frac{1}{2}$} - c \right) }}=1 \ .
\eeq This equation appears only
at second order in $\frac1p$ because we have
summed the equations with a weight depending on $p$ in order to have
the simplest possible expressions. This procedure mixes the various
orders of the expansion.

\subsection{Phase transitions\label{Phasediag}}

The problem of the existence of solutions to eqs.~(\ref{qTC1})
and~(\ref{qTC2}) is most easily addressed by introducing a new set
of variables that are centered around the point $\frac12$, given by
the K\"ahler modulus $t=\frac A{2\,p}\,(p-2)$ and the new endpoint
parameters \beq
\hat b=d^\prime-b^\prime=\mbox{$\frac{2\,t}{p-2}\,
\left(\frac{1}{2}-b\right)$} \ , \quad
\hat c=d^\prime-c^\prime=\mbox{$\frac{2\,t}{p-2}\,\left(
\frac{1}{2}-c\right)$} \ . \eeq In terms of these
parameters eq.~(\ref{qTC1}) takes the form \beq-p\,\log\left(
\mbox{$\frac{\e^{{\hat{b}}/{2}} + \e^{{\hat{c}}/{2}}}{2}$}\right) +
\log\left(\mbox{$\frac{\left(\sqrt{\e^{\hat{b}}-1}+
\sqrt{\e^{\hat{c}}-1}~\right)^2}{\e^{\hat{b}}-\e^{\hat{c}}}$}\right)=-\frac{
t}{2}~\frac{p+2}{p-2} \ , \label{qTC1new}\eeq while eq.~(\ref{qTC2}) becomes
 \beq
p\,\log\left(\mbox{$\frac{\e^{-{\hat b}/{2}} +\e^{-{\hat
c}/{2}}}{2}$}\right)+ \log\left(\mbox{$\frac{\left(\sqrt{1-\e^{-\hat{b}}}+
\sqrt{1-\e^{-\hat{c}}}~\right)^2}{\e^{-\hat{c}}-\e^{-\hat{b}}}$}\right)
=\frac{{t}}{2} \ .
\label{qTC2new}\eeq
In principle one should now fix the
coupling constant $t$ and solve for the endpoints $\hat b$ and
$\hat c$. However, this approach is not the most practical one for a
numerical analysis. Thus we choose instead to fix $x=\e^{-{\hat
    c}/{2}}\in[0,1]$ and determine the corresponding variables
$y=\e^{-{\hat b}/{2}}\in[0,1]$ and $t$. From eq.~(\ref{qTC2new}) we
have \beq \label{rippo1} t=2\,p\,\log\left(\mbox{$\frac{x + y}2$}\right)
+4\,\log\left(\mbox{$\frac{\sqrt{1 - x^2} +\sqrt{1
- y^2}}{\sqrt{x^2 - y^2}}$}~\right) \ ,  \eeq
and substituting this into eq.~(\ref{qTC1new}) yields \beq \label{rippo2}
T(x,y;p):=-1+\frac{4^{\frac{p}{ p-2}}\,x\,y\,{\left( x^2 - y^2 \right)
}^{\frac{4}{\left( p-2 \right) \,p}}\,
     {\left(x\,{\sqrt{1 - y^2}} +y\, {\sqrt{1 - x^2}}~
\right) }^{\frac{2}{p}}}{{\left( x + y \right) }^{\frac{2\,p}{ p-2}}\,
     {\left( {\sqrt{1 - x^2}} + {\sqrt{1 - y^2}}~ \right)
     }^{\frac{2\,\left( 2 + p \right) }{\left(  p-2 \right) \,p}}}=0 \
   .
\eeq Once we have fixed the geometrical datum $p$ and given our
choice for $x$, the algebraic equation (\ref{rippo2})
determines $y$ and eq. (\ref{rippo1}) then gives the K\"ahler modulus
$t$.

\begin{figure}[hbt]
\begin{center}
\epsfxsize=3.0 in \epsfysize=2.0 in \epsfbox{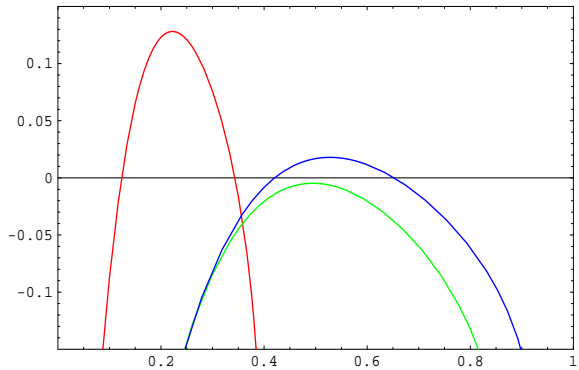}
\end{center}
\caption{Plot of $T(x,y;p)$ as a
  function of $y$ for $p=7$ and $x=0.4\in I_1$ (left curve),
  $x=0.89\in I_2$ (center curve) and $x=0.95\in I_3$ (right curve).}
\label{fundamentaeqchiral}\end{figure}

\begin{figure}[hbt]
\begin{center}
\epsfxsize=2.5 in \epsfysize=1.5 in \epsfbox{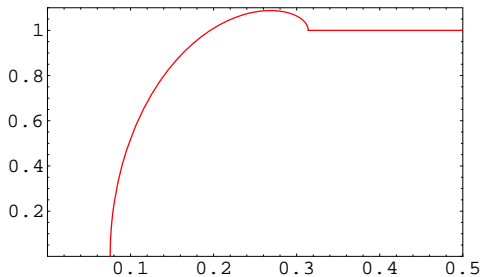}
\end{center}
\caption{Unphysical solution $\rho(s)$ versus $s$ for $x=0.4\in
  I_1$.}
\label{unphysicalrho}\end{figure}

When we take into account the constraint $y\le x$ (i.e. $c\ge b$)
eq.~(\ref{rippo2}) does not always admit a solution for a given
$p$. Generally, the interval $[0,1]=I_1\amalg I_2\amalg I_3$
decomposes into three disjoint, connected subintervals in the
following manner. For $x\in I_1$ there are two solutions $y$ of
eq.~(\ref{rippo2}), for $x\in I_2$ there are no solutions, and for
$x\in I_3$ there are again two solutions. The sizes of the of
these subintervals depend on $p$. The region $I_1$ increases as
$p$ grows, while $I_2$ and $I_3$ decrease very rapidly. A
representative example of this behaviour is depicted in
Fig.~\ref{fundamentaeqchiral}.
We also have to impose the additional
constraint on the spectral distribution that $\rho\leq1$. This requirement
immediately rejects the solution in the interval $I_1$ for which the
density has the form depicted in Fig.~\ref{unphysicalrho}.

\begin{figure}[hbt]
\begin{center}
\epsfxsize=6.0 in \epsfysize=4.4 in \epsfbox{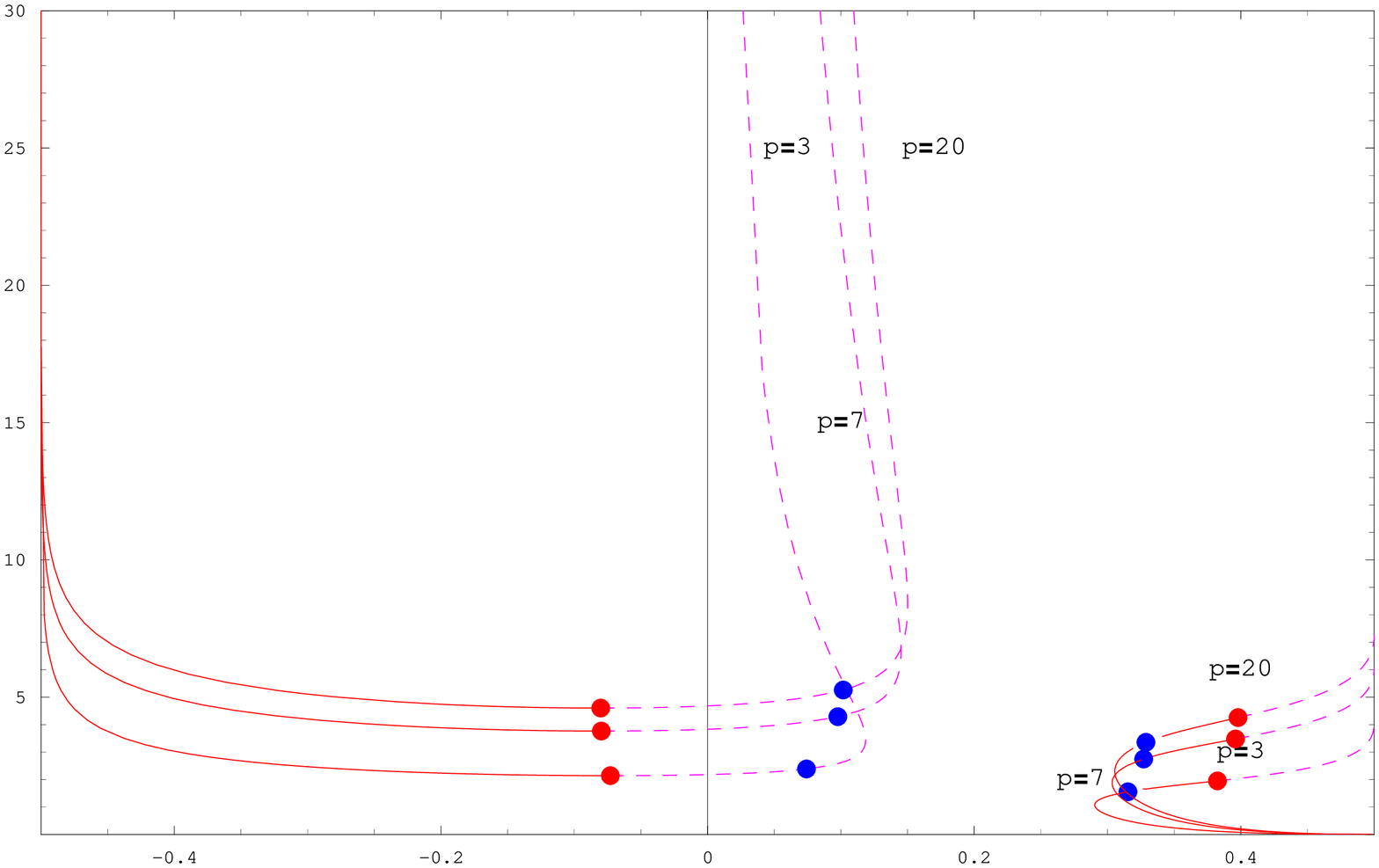}
\end{center}
\caption{Behaviour of $t$ as a function of $c$ for $p=3,7,20$. The
  dashed parts of the curves indicate the unphysical regions having $\rho>1$.
The dots separating a dashed line from a solid line represent the
critical points of a phase transition. The other dots represent
diramation points for the solutions of the equation $T(x,y;p)=0$
for each branch. } \label{cuppa}\end{figure}

A clearer picture of the situation is obtained if we draw the phase
diagram in the $(c,t)$-plane for various values of the geometrical
parameter $p$ (Fig.~\ref{cuppa}).
 The qualitative
behaviour is very similar to that of ordinary chiral QCD$_2$
\cite{Crescimanno:1994eg}. The lines coming from the region of
large $t$ all represent physical solutions up to a certain
critical value $t^+_{\rm
  c}(p)$. This means that the
one-cut solution allows us to explore the region with large K\"ahler
parameter. When the lines reach and go below
$t=t^+_{\rm c}(p)$, the distribution function $\rho$ becomes larger than
$1$ and a phase transition occurs. Rather remarkably, the value $t^+_{\rm
  c}(p)$ is very close to the value of the K\"ahler modulus that
triggers the phase transition in the full coupled $q$-deformed
gauge theory \cite{Caporaso:2005ta,Arsiwalla:2005jb,Jafferis:2005jd}. At
this point no physical solution exists until we reach a second
critical point $t^-_{\rm c}(p)$. This point connects the line
coming from the region of small $t$. The distribution function
$\rho$ is again smaller than $1$ for $0<t<t^-_{\rm c}(p)$ and a
second phase transition occurs at $t=t^-_{\rm c}(p)$. We conclude
that our one-cut solution describes the large $N$ gauge theory in
two distinct phases, one which covers the large values of the
K\"ahler parameter $t$ and the other describing the small values
of $t$. To connect these two phases, one would have to construct
an appropriate two-cut solution in the intermediate region.

\subsection{Topological strings in the strong-coupling
  phase\label{Topstrong}}

We expect that the topological string theory on $X_p$ will emerge from
the $q$-deformed gauge dynamics when $t$ is large, i.e. for $t>t_{\rm
  c}^+(p)$ where the one-cut solution constructed above is valid. We
therefore seek a consistent expansion of the solution of the
saddle-point equation for large values of the coupling constant $t$. For this,
we assume that the endpoints $b$ and $c$ are finite in the limit $t\to\infty$,
or equivalently that $\hat b$ and $\hat c$ diverge at most linearly in
$t$. Then eqs.~(\ref{qTC1}) and~(\ref{qTC2}) reduce for large $t$ to
\beq \hat b=\hat c=\mbox{$\frac{2\,t}{p-2}$}=2\,\hat t \ , \eeq or equivalently
$b=c=-\frac12$. This means that at strong-coupling the distribution
function $\rho(z)$ tends to become flat and symmetric about the
origin. The particular form of the saddle point
equation (\ref{piou}) suggests that corrections to this result are
exponentially suppressed in $t$. Thus we look for a solution
of the form \bea \hat b&=&2\,\hat t+\sum_{n=1}^\infty\, r_n~\tau^{n}
\ , \nonumber\\ \hat c&=&2\,\hat t+\sum_{n=1}^\infty\,s_n~\tau^{n}
\eea where $\tau:=\e^{-\xi\,t/2}$. Imposing consistency of our ansatz
fixes the parameter $\xi=1$.

Proceeding iteratively, we arrive at \bea \hat
b&=&2\,\hat t+\left[2 -
  2{~\e^{-2\,\hat t}}\,\right]~\e^{-t/2}+
\left[1 - p + {2\,p}{~\e^{-2\,\hat t}}-
(1+{p}){~\e^{-4\,\hat t}} \,\right]~\e^{-t}\nonumber\\ && +\,
\left[\mbox{$\frac{2}{3}$}-2\,p+p^2 +
 \left({2\,p}-3\,p^2\right){~\e^{-2\,\hat t}}+\left( {2\,p}+3\,p^2\right){~
\e^{-4\,\hat t}} \right.\nonumber\\ && -\left.\left(
\mbox{$\frac{2}{3}$}+2\,p+p^2\right)~\e^{-6\,\hat t}\,
 \right]~\e^{-{3\,t }/{2}}+O\left(\e^{-2\,t}\right) \ , \nonumber\\[5pt]
\hat c&=&2\,\hat t-\left[2 -
2{~\e^{-2\,\hat t}}\,\right]~\e^{-t/2}+ \left[1  - p +
{2\,p}{~\e^{-2\,\hat t}}- (1+{p}){~\e^{-4\,\hat t}}\,
\right]~\e^{-t}\nonumber\\ && -\,
\left[\mbox{$\frac{2}{3}$}-2\,p+p^2 +
 \left({2\,p}-3\,p^2\right){~\e^{-2\,\hat t}}+\left( {2\,p}+3\,p^2\right){~
\e^{-4\,\hat t}} \right.\nonumber\\ && -\left. \left(
\mbox{$\frac{2}{3}$}+2\,p+p^2\right)~\e^{-6\,\hat t}\,
 \right]~\e^{-{3\,t }/{2}}+O\left(\e^{-2\,t}\right) \ .
\eea Note that the corrections are order by order
polynomials in $\e^{-2\,\hat t}$. This is very different from
what happens in the full coupled theory where the corrections are
given by infinite series. If we identify the power of $\e^{-{t}/{2}}$
with the winding number of the topological string expansion, then the
behaviour of our solutions is exactly that expected from string
theory. We recall from Section~\ref{Chiralexp} that the large $N$ expansion of
the partition function was organized in exactly the same way.

Our next goal is to compute the chiral partition function for large
values of the K\"ahler parameter $t$. For this, we will calculate the
derivative of the free energy with respect to the area $A$ at fixed
$\hat t=\frac{g_s N}2$. We have \cite{Caporaso:2005ta}
\beq \frac{\partial\mathcal{F}_0\big(A\,,\,\hat t\,;\,p\big)}{\partial
A}=\frac{1}{2}\,\int^{{1}/{2}}_{b}\dd x~x^2\,\rho(x)=
\frac{1}{2}\,\int^{c}_{b}\dd x~x^2\,\rho(x)+
\frac{1}{2}\,\left(\frac{1}{24}-\frac{c^3}{3}\right) \eeq and so we
need to compute the integral \beq {\cal I}:=\frac{1}{2}\,\int^{c}_{b}\dd x~x^2
\,\rho(x) \ . \eeq We change variable
$x=p\,\log(\e^{-2\,A\,c/p^2 }\,u )/A$ to get \beq {\cal I}=\frac{p^2}{2\,
A^2}\,\int^{\e^{c^\prime}}_{\e^{b^\prime}}\dd u~\log^2\big(\e^{-2\,A\,
c/p^2 }\,u \big)\,\varrho(u) \ . \eeq To express the result in terms
of the endpoint parameters $\hat b$ and $\hat c$, we set $u=s~
\e^{d^\prime}$ and use the explicit solution (\ref{1cutexpl}) to obtain
\bea {\cal I}&=&\frac{p^4}{2 \pi\,A^3}\,\int^{\e^{-\hat c}}_{\e^{-\hat b}}
\,\frac{\dd s}{s}~\log^2\big(\e^{{A}/{2\,p} }\,s\big)\,\biggl[
\arctan\left(\mbox{$\sqrt{\frac{\e^{-\hat c}-s}{s-\e^{-\hat
b}}}$}~\right)-\arctan\left(\mbox{$\e^{({\hat c}-{\hat
b})/{2}}\,\sqrt{\frac{\e^{-\hat c}-s}{s-\e^{-\hat b
}}}$}~\right)\biggr.\nonumber\\
&&\ \ \ \ \ \ \ \ \ \qquad\qquad\qquad\qquad\qquad\qquad
+\biggl.\frac{2 }{ p }\,
\arctan\left(\mbox{$\sqrt{\frac{s-\e^{-\hat b}}{\e^{-\hat
c}-s}\,\frac{1-\e^{-\hat c}}{1-\e^{-\hat b}}}$}~\right)\biggr] \ . \eea
To simplify the analysis, it is convenient to integrate by
parts to get \bea {\cal I}&=&
\frac{c^3}{6}-\frac{1}{96\pi\,A^3\,
    {\sqrt{1 - \e^{-\hat c}}}}\,\int_{\e^{-\hat b}}^{\e^{-\hat c}}\,
\frac{\dd s }{s\,(s-1)\,
   {\sqrt{(\e^{-\hat c} - s)(s-\e^{-\hat b} )}}}\nonumber\\ &&\times\,
    \biggl[\e^{-{(\hat b+\hat c)}/{2}}\,{\sqrt{1 - \e^{-\hat c}}}\,
         p\,\left( 1-s \right)-
      2~\e^{-\hat c}\,{\sqrt{1 - \e^{-\hat b}}}\,s
   \biggr. \nonumber\\ &&+\biggl.\left( 2\,{\sqrt{1 - \e^{-\hat b}}} +
         {\sqrt{1 - \e^{-\hat c}}}\,p\,\left( s -1\right)
         \right) \,s \biggr] \,
    {\bigl( A + 2\,p\,\log (s) \bigr) }^3 \ .
\eea

Analyzing the behaviour of this integral directly is hampered by the
fact that we are exploring the singular region where the two endpoints
begin to coincide. We will therefore perform a change
of variable that makes the region of integration independent of $t$.
The transformation
 \beq s=\mbox{$\frac12$}\,\left[\e^{-\hat b} + \e^{-\hat c}-
\bigl( \e^{-\hat c} -\e^{-\hat b}\bigr) \,\cos (\theta )\right]
\eeq maps the integration domain onto $\theta\in[0,\pi]$. It also
eliminates the square root in the denominator of the integrand in
$\cal I$ and we arrive at
\bea {\cal I}&=&\frac{c^3}{6}-\frac{1}{48\pi\,A^3\,{\sqrt{1
- \e^{-{\hat c}}}}}\,\int_0^\pi\,\frac{\dd\theta}
    {\e^{{\hat b}} + \e^{{\hat c}} -
      2~\e^{{\hat b} + {\hat c}} +
      \bigl(\e^{{\hat c}}-\e^{{\hat b}}
         \bigr) \,\cos (\theta )}\nonumber\\
&&\times\,\frac{1}{
      \e^{{\hat b}} +
      \e^{{\hat c}} +
      \bigl(\e^{{\hat c}} -\e^{{\hat b}}
         \bigr) \,\cos (\theta )}\,\biggl[ 2~
       \e^{{{\hat b}}}\,
       {\sqrt{1 - \e^{-{\hat b}}}}\,
       \left(\e^{{\hat b}}+\e^{{\hat c}}-
         \bigl( \e^{{\hat b}} -
            \e^{{\hat c}} \bigr) \,\cos (\theta )
         \right)\biggr.\nonumber\\
&&-~\e^{{{(\hat b+\hat c)}}/{2}}\,{{\sqrt{1 -
             \e^{-{\hat c}}}}\,p\,
         \left( \e^{{\hat b}} + \e^{{\hat c}} -
           2~\e^{{\hat b} + {\hat c}} +
           \bigl(\e^{{\hat c}}-\e^{{\hat b}}
              \bigr) \,\cos (\theta )
           \right) }{} \nonumber\\ && +\,
       \left( \e^{{\hat b}} + \e^{{\hat c}} +
         \bigl( \e^{{\hat c}}-\e^{{\hat b}}
             \bigr)\,\cos (\theta )
         \right)\nonumber\\ && \times\biggl.\left( 2~\e^{{{\hat b}+\hat c}}\,
          {\sqrt{1 - \e^{-{\hat b}}}} +\left.
         \frac{
            {\sqrt{1 - \e^{-{\hat c}}}}\,p
            }{2}\,\left( \e^{{\hat b}} +
              \e^{{\hat c}} -
              2~\e^{{\hat b} + {\hat c}} +
              \bigl( \e^{{\hat c}}-\e^{{\hat b}}
                  \bigr) \,
               \cos (\theta ) \right)\right)
      \biggr]\right.\nonumber\\
&&\times\,{\left[A+
        2\,p\,\log\left(\mbox{$\frac{\e^
              {-{\hat b} - {\hat c}}\,
             \left( \e^{{\hat b}} +
               \e^{{\hat c}}-
               (\e^{{\hat b}}-
                  \e^{{\hat c}}) \,
                \cos (\theta ) \right) }{2}$}\right) \right]}^3  \ .
\eea

Expanding $\cal I$ as a series in $\e^{-t}$ and integrating
over $\theta$, we arrive finally at
\beq \frac{\partial {\cal F}_0\big(A\,,\,\hat t\,;\,p\big)}{\partial
A}=\frac{{\bigl(1 - \e^{-2\,\hat t}\,\bigr) }^2\,{\left( p-2
\right) }^2}{8\,t^2}~\sum_{n=1}^\infty\,\e^{-n\,t}~G_n\big(t\,,\,\hat
t\,;\,p\big) \ , \label{Iexp}\eeq
where the first three contributions are given by\bea
G_1&=&1 \ , \nonumber\\[5pt]
G_2&=&\frac{\e^{-{4\,\hat t}}}4\,\left[1 + 2\,p\,\left( 2 +
p \right)  + \e^{4\,\hat t}\,\bigl( 1 + 2\,\left( p-2 \right)
\,p \bigr)  +
     2~\e^{2\,\hat t}\,\left( 1 - 2\,p^2 \right)\right]\nonumber\\
  &&-\,\frac{\e^{-4\,\hat t} }
   {2\,t}\,\left[\bigl(1 -\e^{2\,\hat t}\,\bigr) \,\bigl(1 +
\e^{2\,\hat t}\,\left( p-1\right)  - p \bigr) \,\left( p-2
\right)\right] \ ,
\\[5pt]
G_3&=&\mbox{$\frac{1}{9} - p + \frac{5}{2}\,p^2 - 2\,p^3 + \frac{1}{2}\,p^4 +
\e^{-6\,\hat t}\,{\left(\frac{2}{9} + p - p^2 - 4\,p^3 -
2\,p^4\right)}{}$}
\nonumber\\
&&\mbox{$+~\e^{-2\,\hat t}\,{\left(\frac{2}{9} - p - p^2 + 4\,p^3 -
2\,p^4\right)}{} +\e^{-8\,\hat t}\,
  {\left(\frac{1}{9} + p + \frac{5}{2}\,p^2 + 2\,p^3 +
      \frac{1}{2}\,p^4\right)} $} \nonumber\\
&&\mbox{$+~\e^{-4\,\hat t} {\left(\frac{1}{3} - 3\,p^2 +
3\,p^4\right)}{}$} -
  \frac{1}{t}\,\Bigl[\mbox{$\frac{2}{3}-\frac{11}{3}\,p+
\frac{17}{3}\,p^2 -\frac{10}{3}\,p^3 +\frac{2}{3}\,p^4$}\Bigr.
\nonumber\\
&&+~\e^{-4\,\hat t}\,{\left(4\,p -2\,p^2 - 8\,p^3 +
4\,p^4\right)}{} -{\e^{-{8\,\hat t}}}\,
    {\left(\mbox{$\frac{2}{3} + 3\,p + \frac{7}{3}\,p^2 -
          \frac{2}{3}\,p^3
 - \frac{2}{3}\,p^4$}\right)}\nonumber\\
&&\left.+~\e^{-2\,\hat t}\,\mbox{$
     {\left(\frac{2}{3}+p-\frac{26}{3}\,p^2 + \frac{28}{3}\,p^3 -
 \frac{8\,p^4}{3}\right)} -{\e^{-6\,\hat t}}\,
     {\left(\frac{2}{3} - \frac{5}{3}\,p - \frac{22}{3}\,p^2
 - \frac{4}{3}\,p^3 + \frac{8}{3}\,p^4\right)}$}\right] \ . \nonumber
\eea
To obtain the free energy we now have to integrate over the area
$A$. For this, we express $p$ and $t$ in terms of $A$ and $\hat t$ as
$p=\frac{A}{2\,\hat t}$ and $t= \frac{A}{2}-2\,\hat t$ to write the
expansion (\ref{Iexp}) in terms of
\beq
\Gamma_n\big(\hat
t\,;\,A\big):=\frac{1}{8\,\hat t^{\,2}}\,\left(1-\e^{-2\,\hat t}\,\right)^2~
\e^{-n\,\big(\frac A2-2\,\hat t\big)}~G_n\big(\,\mbox{$\frac A2-2\,\hat
  t\,,\,\hat t\,;\,
\frac A{2\,\hat t}$}\,\big) \ . \eeq
For the first three terms one has
\bea
\Gamma_1&=&\frac{\e^{-\frac{A}{2} - 2\,{\hat t}}\,{\bigl
(1-\e^{2\,{\hat t}}\,\bigr) }^2}{8\,{{\hat t}}^{\,2}} \ , \nonumber\\[5pt]
\Gamma_2&=&\frac{\e^{-A - 4\,{\hat t}}\,{\bigl(1 - \e^{2\,{\hat
t}}\,\bigr) }^2\,
    }
    {64\,{{\hat t}}^{\,4}}\,\left[A^2\,{\left(1 -
\e^{2\,{\hat t}} \right) }^2 - 2\,A\,\left(1 -\e^{2\,{\hat t}} \right) \,
       \left(1 - 2\,{\hat t} -\e^{2\,{\hat t}}\,\left( 1 +2\,{\hat t}\,
\right)  \right)\right.\nonumber\\ &&-\left.4\,
      \left( 1 + \e^{2\,{\hat t}} \right) \,{\hat t}\,
\left(2 - {\hat t} -\e^{2\,{\hat t}}\,\left(2 + {\hat t}\,
\right)  \right)  \right] \ , \nonumber\\[5pt]
\Gamma_3&=& \frac{\e^{-\frac{3}2\, \left(A + 4 \,{\hat t}\,
\right) }\,{\bigl(1-\e^{2\,{\hat t}}\,\bigr) }^2
}{2304\,{{\hat t}}^{\,6}}\,
\left[\,9\,A^4\,{\left(1-\e^{2\,{\hat t}}\right) }^4\right.
\nonumber\\ && -\,24\,A^3\,
{\left(1 -\e^{2\,{\hat t}} \right) }^3\,
\left(1 - 3\,{\hat t} -\e^{2\,{\hat t}}\,
\left(1 + 3\,{\hat t}\,\right)  \right)\nonumber\\
&& +\,36\,A^2\,{\left(1 -\e^{2\,{\hat t}} \right) }^2\,{\hat t}\,
       \left(4 - 5\,{\hat t} - 8\,{\hat t}~\e^{2\,{\hat t}}-
\e^{4\,{\hat t}}\,\left( 4 + 5\,{\hat t}\,\right)  \right)
\nonumber\\ &&  -\,48\,A\,\left(1 -\e^{{\hat t}} \right)\,{{\hat t}}^{\,2}\,
       \left(5 +3~\e^{2\,{\hat t}}\,\left(1 -2\,{\hat t}\,\right)-3\,
{\hat t} -3~\e^{4\,{\hat t}}\,\left( 1 + 2\,{\hat t}\,
\right)  -\e^{6\,{\hat t}}\,\left( 5 + 3\,{\hat t}\,\right)  \right)
\nonumber\\ && -\biggl.32\,{{\hat t}}^{\,3}\,
\left( 1 + \e^{2\,{\hat t}} + \e^{4\,{\hat t}} \right)\,
       \left(3-{\hat t}-{\hat t}~\e^{2\,{\hat t}} -\e^{4\,{\hat t}}\,
\left( 3 + 2\,{\hat t}\,\right)  \right)\biggr] \ .
\eea

These expressions are now easily integrated over $A$ and we
arrive at the free energy components
\beq
F_n\big(\hat t\,;\,p\big):= 4{\hat t}^2\left(1-\e^{-2\,\hat
  t}\,\right)^{-2}~\e^{n\,\big(\frac A2-2\,\hat t\big)}~
\left.\biggl[~\int\dd A~\Gamma_n\big(\hat
t\,;\,A\big)\biggr]\right|_{A=2\,p\,\hat t} \ .
\eeq
For the first three contributions we
find \bea
F_1&=&1 \ , \nonumber\\[5pt]
F_2&=&\frac{\e^{-4\,\hat t}}
  {8}\,\left[1 + 4\,p + 2\,p^2 + \e^{2\,\hat t}\,\left( 2 - 4\,p^2
    \right)+\e^{4\,\hat t}\,\left( 1 - 4\,p + 2\,p^2 \right)
  \right] \ , \nonumber\\[5pt]
F_3&=&\frac{\e^{-8\,\hat t}}{54}\,\left[2 + 18\,p + 45\,p^2 +
36\,p^3 + 9\,p^4 + 6~\e^{4\,\hat t}\,\left( 1 - 9\,p^2 +
9\,p^4 \right)\right.\nonumber\\
&&+  \left.  ~\e^{8\,\hat t}\,\left( 2 - 18\,p + 45\,p^2 -
36\,p^3 + 9\,p^4 \right)+
    2~\e^{6\,\hat t}\,\left(2 - 9\,p - 9\,p^2 + 36\,p^3 - 18\,p^4
\right)\right.\nonumber\\ &&+\left.
    2~\e^{2\,\hat t}\,\left(2 + 9\,p - 9\,p^2 - 36\,p^3 - 18\,p^4
\right)\right] \ . \eea
These expressions coincide with the ones obtained in (\ref{frees}) for
the topological string theory on $\hat X_p$.

\section{Analytic properties of the topological string perturbation
  series\label{stringseries}}

Motivated by the saddle-point analysis of the phase structure of
the chiral gauge theory, in this section we shall investigate the
convergence properties of the perturbative topological string
expansion of the partition function. In ordinary chiral Yang-Mills
theory on $S^2$ the large $N$ phase transition can be analysed
from the string theory perspective~\cite{Taylor:1994zm}. Analytic
and numerical results indicate that the string perturbation series
has a finite radius of convergence which coincides with the
critical points. In the Gross-Taylor string language, the phase
transition is driven by the entropy of branch point singularities
of the string covering maps \cite{Taylor:1994zm}. Here we will
perform an analogous investigation for the $q$-deformed chiral
gauge theory and show that a similar picture emerges, thereby
supporting the results of the previous section.

\subsection{Genus zero\label{Genus0}}

Recall that the genus zero free energy has the form  \beq
\label{gw} {\cal F}_0\big(t\,;\,p\big)=
\sum_{n=1}^{\infty}\,\e^{-n\,t }~\sum_{k=0}^{2n}
\,\e^{-\frac{2\,k\,t}{p-2} }~{\rm N}^0{}^{~}_{n,k}(p)
\eeq where ${\rm N}^0{}^{~}_{n,k}(p)$ is a polynomial in $p$ of
degree $2n-2$. These polynomials are given in eq.~(\ref{frees}) for the
first six degrees, and generally they can be parameterized as
\beq {\rm N}^0{}^{~}_{n,k}(p)=\sum_{m=0}^{2n-2}\,{\rm C}^0{}^{~}_m(n,k)~p^m \
. \label{gwpar}\eeq To study the convergence properties of the
series (\ref{gw}), we investigate its asymptotic large $n$
behaviour. The dominant contributions from the genus zero Gromov-Witten
invariants (\ref{gwpar}) are the highest degree monomials
\beq \label{approx} \lim_{n\to\infty}\,{\rm
  N}^0{}^{~}_{n,k}(p)={\rm C}^0{}^{~}_{2n-2}(n,k)~ p ^{2n-2} \ . \eeq We are
thereby led to analyse the series \beq {\cal
  F}^\infty_0\big(t\,;\,p\big)=\sum_{n=1}^{\infty}\,\e^{-n\,t }\,
p^{2n-2}~\sum_{k=0}^{2n}\,{\rm C}^0{}^{~}_{2n-2}(n,k)~\e^{-\frac{2\,k\,
t}{p-2}} \ . \label{largenseries}\eeq

From a direct inspection of the first six orders in
eq.~(\ref{frees}), we conjecture that the combinatorial coefficients in
(\ref{approx}) are given by \beq
{\rm C}^0{}^{~}_{2n-2}(n,k)= (-1)^k\,\frac{n^{n-3}}{n!}\,{2n\choose k} \ .
\label{guess}\eeq We will motivate this conjecture below. It implies
that the sum over $k$ in (\ref{largenseries}) can be carried out
explicitly and the expansion written as
\beq {\cal F}^\infty_0\big(t\,;\,p\big)=\sum_{n=1}^{\infty}\,
 \frac{n^{n-3}}{n!}\,p^{2n-2}\,\left(1-\e^{-\frac{2\,t}{p-2} }
 \right)^{2n}~\e^{-n\,t } \ . \eeq
\noindent The radius of convergence of this
series is easily determined. By substituting the Stirling
approximation
\beq
n! ~\stackrel{n\to\infty}{\longrightarrow}~
\sqrt{2 \pi\,n }~n^n~\e^{-n }
\eeq
we find that the large $n$ behaviour of the free energy converges for
\beq \label{radius}
  p^{2}\,\left(1- \e^{-\frac{2\,t}{p-2} }  \right)^{2}~\e^{- (t-1) }
  \leq 1 \ . \eeq

We can perform a simple check of this bound by looking at the limit in
which the $q$-deformed gauge theory reduces to ordinary Yang-Mills
theory. Rewriting eq.~(\ref{radius}) in terms of the QCD$_2$ area
parameter $A=N\,g_sp$ gives \beq
  p^{2}\,\left(1-\e^{-{  A}/{p} }  \right)^{2}~\e^{-\big(\frac
    A2-\frac Ap-1\big) } \leq 1 \ , \eeq
and in the limit  $ p \rightarrow \infty$ with $A$ fixed we find
\beq
  A^2~\e^{- \big(\frac A2-1\big) } \leq 1 \ . \label{QCD2conv}\eeq
This coincides with the convergence bound derived in
\cite{Taylor:1994zm}.

To understand better the
approximation (\ref{approx}) and the extrapolation of the coefficients
(\ref{guess}) to all orders, we need to take a closer look at the
mechanism by which we recover the chiral QCD$_2$ string perturbation series
from eq.~(\ref{gw}). For this, let us reinsert the string coupling
constant $g_s$ as in eq.~(\ref{gsincl}) to write
\beq \label{gsapproxrewrite}{\cal F}_0(A;p)=
\frac{N^2 p^2}{A^2}\,\sum_{n=1}^{\infty}\,\e^{-n\,A /2 }~
\sum_{k=0}^{2n}\,\e^{ (n-k)\,{  A}/{p} }~{\rm N}^0{}^{~}_{n,k}(p)
\, \eeq
and expand the free energy in the limit $p\to\infty$ as \beq
{\cal F}_0(A;p)=\frac{N^2 p^2}{A^2}\,\sum_{n=1}^{\infty}\,\e^{-n\,A /2
}~\sum_{k=0}^{2n}~\sum_{m=0}^{2n-2}\,
{\rm C}^0{}^{~}_m(n,k)~p^m~\sum_{l=0}^{\infty}\,\frac{1}{l!}\,
\left(\mbox{$\frac{A}{p}$}\,(n-k) \right)^l \ .
\label{calF0pinfty}\eeq
In the undeformed limit, the only contribution
to the leading area term $A^{2n-2}$ comes
from the combination (\ref{approx}). The assumption
that eq.~(\ref{guess}) is valid at all orders implies that the
coefficient of $A^{2n-2}$ is given by \bea \label{undefcontr}
\sum_{k=0}^{2n}\,\frac{\left( n-k \right)^{2n }}{(2n)!}~
{\rm C}^0{}^{~}_{2n-2}(n,k)= \sum_{k=0}^{2n}\,\frac{ (-1)^k}{(2n)!}\,\left( n-k
\right)^{2n }\,\frac{n^{n-3}}{n!}\,{2n\choose k}=\frac{n^{n-3}}{n!} \
. \eea
In~\cite{Crescimanno:1994eg} it was shown that
this is exactly the pertinent coefficient of
$A^{2n-2}$. It is equal to the number of (topological classes of)
holomorphic maps $\PP^1\to\PP^1$ with $2n-2$ simple branch point
singularities. In the given limit the free
energy (\ref{calF0pinfty}) thus reduces to
\bea \label{conv}
\Phi_0^\infty(A)=N^2\,\sum_{n=1}^{\infty}\,\frac{n^{n-3}}{n!}~A^{2n-2}~
\e^{-n\,A /2 } \ . \eea
In the Gross-Taylor string description, this is the contribution to
the chiral QCD$_2$ free energy coming from covering maps with only
branch points and no $\Omega$-point singularities. Our conjecture
(\ref{guess}) thereby passes a highly non-trivial check.

Subleading terms in the limit $ p\rightarrow \infty$ with $A$ fixed
must then reproduce the full undeformed chiral
free energy. In ordinary chiral Yang-Mills theory,
the upper radius of convergence of the string perturbation series
(\ref{conv}) can be estimated to be $A_{\rm c}^+\simeq 11.9$, while the one
evaluated from the saddle-point equation for the full chiral free
energy is approximately $10.189$ \cite{Crescimanno:1994eg}. The string
estimate gives a value which is $0.2$ times larger than the actual
one. In the $q$-deformed gauge theory one can also compare the result
coming from an estimate of the radius of convergence of the string
perturbation series with the numerical evaluation of the saddle-point
equation from Section~\ref{Phasediag}, and with the exact critical point
of the full coupled gauge theory. The results are summarized in
Table~\ref{tcritcompare}.
\begin{table}[hbt]
\begin{center}
\begin{tabular}{|l|l|l|l|}
  \hline
  $p$ & ${\hat t_{\rm nc}(p)=\frac{2\, t_{\rm nc}(p)}{p-2}}$ & $
  {\hat t^{\,{\rm num}}_{\rm c}(p)=\frac{2\,t^{\,{\rm num}}_{\rm c}(p)}{p-2}}$ & ${
  \hat t^{\,{\rm str}}_{\rm c}(p)= \frac{2\, t^{\,{\rm str}}_{\rm c}(p)}{p-2}}$
  \\ \hline\hline
  1000 & $0.99\times 10^{-2}$ & $1.02\times 10^{-2}$ & $1.19\times 10^{-2}$
  \\ \hline
  300 & $3.29\times 10^{-2}$ & $3.40\times 10^{-2}$ & $3.97\times 10^{-2}$
  \\ \hline
  100 & $0.99\times 10^{-1}$ & $1.02\times 10^{-1}$ & $1.19\times 10^{-1}$
  \\ \hline
  30 & $3.30\times 10^{-1} $ & $3.40 \times 10^{-1} $ & $3.98 \times 10^{-1} $
  \\ \hline
  10 & $1.00$ & $1.04$ & $1.23$
  \\ \hline
  7 & $1.46$ & $1.51$ & $1.81$
  \\ \hline
  5 & $2.12$ & $2.19$ & $2.72$
  \\ \hline\hline
  3 & $4.16$ & $4.28$ & $6.39$
  \\ \hline
  2.1 & $10.9$ & $11.1$ & $49.7$
  \\ \hline
  2.01 & $19.5$ & $19.8$ & $479$
  \\ \hline
  2.001 & $28.6$ & $28.9$ & $4770$
  \\ \hline
\end{tabular}
\end{center}
\caption{The non-chiral critical coupling constant $\hat{t}_{\rm nc}(p)$
is obtained from the full coupled gauge
theory~\cite{Caporaso:2005ta,Arsiwalla:2005jb,Jafferis:2005jd}. The chiral
critical coupling $\hat{t}^{\,{\rm num}}_{\rm c}(p)$ is the larger of the two
critical points obtained from the two endpoint equations of the
previous section \textit{plus} the additional equation
$\rho'(c)=0$ signalling the boundary of the physical region. The
critical point $\hat{t}^{\,{\rm str}}_{\rm c}(p)$ is evaluated
from the topological string perturbation series as the larger value of
the two solutions to the estimate for the radius of convergence. We have
normalized the variables with a factor $2/(p-2)$ because, with this
choice, the equation makes sense for any value of $p$.}
\label{tcritcompare}\end{table}
We see that the difference between the non-chiral and chiral critical
normalized couplings $\hat{t}_{\rm nc}(p)=2\,t_{\rm nc}(p)/(p-2)$ and $
\hat{t}^{\,{\rm num}}_{\rm c}(p)=2\,t^{\,{\rm num}}_{\rm c}(p)/(p-2)$
is roughly constant throughout the range of $p$ and of the order $2-3\times
10^{-3}$. The difference between the string determined and
numerically determined normalized critical couplings is roughly 20\% for $p>3$,
and then starts to grow out of control for smaller values of $p$. We
have also included non-integer values of $p$ to illustrate that
$ \hat{t}_{\rm c}^{\,{\rm num}}(p)$ and $ \hat{t}_{\rm nc}(p)$ both diverge
with the same values as $p\rightarrow 2$, reflecting the absence of a
large $N$ phase transition for~$p\leq2$.

It is interesting to note that in the $k=0$ sector of the chiral
theory where the fiber D-branes are neglected, the closed
topological string amplitude on $X_p$ also has a finite radius of
convergence that can be estimated in the same way. The $k=0$
contribution to the chiral free energy (\ref{largenseries}) is given
by $\sum_{n\geq1}\,\frac{n^{n-3}}{n!}\,p^{2n-2}~\e^{-n\,t }$, which
converges for
\beq
p^{2}~\e^{- (t-1) } \leq 1 \ .
\label{braneless}\eeq
This is very similar to the chiral QCD$_2$ result (\ref{QCD2conv}).

\subsection{Genus one\label{Genus1}}

Before moving on to discuss the implications of these results, it is
natural to ask whether or not the finite radius of convergence of the
perturbative expansion of the free energy is an artifact of the genus
zero approximation to the full string theory. We will now show that
this behaviour persists at higher genera by computing the radius of
convergence of the genus one
partition function. Its generic form is similar to that of the genus
zero case and is given by \beq {\cal F}_{1}\big(t\,;\,p\big)=
\sum_{n=1}^{\infty}\,\e^{-n\,t}~\sum_{k=0}^{2n}\,
\e^{-\frac{2\,k\,t}{p-2}}~{\rm N}^{1}{}^{~}_{n,k}(p) \ ,
\eeq where the genus one Gromov-Witten invariants ${\rm
  N}^1{}^{~}_{n,k}(p)$ are polynomials in the variable $p$ of degree
$2n$. Explicit forms for these polynomials are given for the first
five degrees in Appendix~\ref{Genus1partfn}, and in general they may be
parameterized as \beq {\rm N}^1{}^{~}_{n,k}(p)=
\sum_{m=0}^{2n}\,
{\rm C}^1{}^{~}_{m}(n,k)~p^{m} \ . \eeq At large $n$ the dominant
contributions again come from the maximal degree monomials
 \beq \label{approx1} \lim_{n\to\infty}\,{\rm N}^1{}^{~}_{n,k}(p)=
{\rm C}^1{}^{~}_{2n}(n,k)~p^{2n} \ , \eeq and the radius of
convergence can be estimated from the infinite series \beq
{\cal F}_1^\infty\big(t\,;\,p\big)=
\sum_{n=1}^{\infty}\,\e^{-n\,t }\,p^{2n}~\sum_{k=0}^{2n}\,
{\rm C}^1{}^{~}_{2n}(n,k)~
\e^{-\frac{2\,k\,t}{p-2}} \ . \label{calF1infty}\eeq

Analogously to the genus zero case, by inspection of the first five
orders of the series we conjecture that \beq
{\rm C}^1{}^{~}_{2n}(n,k)=(-1)^k~\frac{R(n)}{(2 n)!}~{2 n \choose k}
\label{guess1} \eeq where \beq R(n)=\frac{(2 n)!}{24\, n!}\,
\left(n^n - n^{n - 1} -
      \sum_{m=2}^n\,(m - 2)!\,{n\choose m}\,n^{n -
        m}\right)=:(2n)!~H(n) \ .
\eeq In the Gross-Taylor description, the combinatorial coefficient
$R(n)$ counts the number of branched coverings of a sphere by a torus
with simple ramification corresponding to the trivial partition
$\vec\mu=(1^n)$~\cite{Vakil}, i.e. with no $\Omega$-point singularities. We can
now perform the sum over $k$ explicitly in (\ref{calF1infty}) to get
\beq {\cal F}_1^\infty\big(t\,;\,p\big)=
\sum_{n=1}^{\infty}\,H(n)\,p^{2n}\,\left(1- \e^{-\frac{2\,t}{p-2} }
\right)^{2n}~\e^{-n\,t } \ . \label{calF1Hn}\eeq
\noindent The radius of convergence of this series can be easily
evaluated by substituting the asymptotic behaviour (see
Appendix~\ref{Genus1partfn})
\beq
\lim_{n\to\infty}\,H(n)=\frac{\e^{n }}{48 n} \ .
\eeq
The free energy (\ref{calF1Hn}) thereby converges when \beq
\label{radius1}
  p^{2}\,\left(1- \e^{-\frac{2\,t}{p-2} }  \right)^{2}~\e^{- (t-1) }
  \leq 1 \ .
\eeq This is identical to the convergence bound obtained at genus zero
in (\ref{radius}), suggesting that the divergence of the topological
string perturbation series is a generic property of the partition
function which holds to all orders in the genus expansion.

\subsection{Phase transition as a Hagedorn transition\label{Hagedorn}}

The results of this section indicate that the perturbative
expansion of the topological string theory under examination has a
finite radius of convergence. In particular, even the closed
topological string amplitude on $X_p$ exhibits this same behaviour
according to eq.~(\ref{braneless}). We immediately see that this
conclusion is incorrect for the cases $p=1,2$. For $p=1$ we cannot
disregard the lower powers of $p$ in estimating the asymptotic
behaviour, and the perturbative contribution to the free energy sums
to a polylogarithm function giving \beq
F^{0}{}^{~}_{{0,0}\atop{0,0}}(t;1)= {\rm
n}^0{}^{~}_{1}(1)~\Li_3\big(\e^{-t}\,\big)\eeq at genus zero
\cite{Gopakumar:1998ki}, as ${\rm n}^0{}^{~}_1(1)$ is the only
non-vanishing Gopakumar-Vafa invariant in this case. The case
$p=2$ is more subtle to handle but since again it leads to just
one non-vanishing integer invariant ${\rm n}^0{}^{~}_1(2)$, the
same formula holds. The cases $p=1,2$ are special both from the
point of view of the saddle-point analysis performed in this paper
and in the full partition function
\cite{Caporaso:2005ta,Arsiwalla:2005jb,Jafferis:2005jd} constructed through
coupled $SU(N)$ representations. Instead, for $p\geq 3$ we expect
our analysis to be correct. The Gopakumar-Vafa invariants in this
case do not seem to vanish after any finite degree, and the
topological string amplitude at genus zero is organized as an
infinite series \beq F^{0}{}^{~}_{{0,0}\atop{0,0}}(t;p)=
\sum_{k=1}^{\infty}\,{\rm
n}^0{}^{~}_k(p)~\Li_3\big(\e^{-k\,t}\,\big) \eeq that can diverge
at some critical value $t=t_{\rm c}(p)$. From
eq.~(\ref{braneless}) we see that the string expansion should
diverge for $t<t_{\rm c}(p)$ with \beq t_{\rm c}(p)=1+2\,\log(p) \
. \eeq

One can now speculate on the meaning of this divergence in the
context of closed topological string theory, and wonder if there
is any method of analytical continuation that would allow the
definition of the theory below the critical radius. Because the
topological string partition function counts BPS bound states of
D0-D2 branes in Type~IIA string theory on $X_p$ weighted by their
BPS energy~\cite{Gopakumar:1998ii,Gopakumar:1998jq}, it is
tempting to propose an interpretation of the divergence as a sort
of Hagedorn transition. The number of BPS states, which are
counted by the Gopakumar-Vafa invariants, grows exponentially
leading to a Hagedorn-like behaviour.

Coming back to the chiral string theory studied in this paper we
see that the string perturbation series converges in two different
regions of ${\rm Re}(t)>0$, in contrast to the non-chiral string
theory. This mimicks the behaviour obtained from
the saddle-point analysis in Section~\ref{Phasediag} (see
Fig.~\ref{cuppa}). We have already noted that the larger critical
point derived from eq.~(\ref{radius}) is in good agreement with
the higher phase transition point. The region of small $t$ instead
does not seem to be well described by the string estimate, as the
big difference between the numerical and the analytical
evaluations of the second critical point shows.

\section{Conclusions}

In this paper we have analysed the large $N$ limit of the chiral sector
 of $q$-deformed Yang-Mills theory on the sphere and its relation
with the topological string, as a part of the project initiated in
\cite{Caporaso:2005ta}. We have found a rich picture where a toric
structure and the geometry of its rationally 
embedded curves emerge from the chiral sector of the gauge theory. Confirming
 the proposal of \cite{Aganagic:2004js,Caporaso:2005ta}, the
 strong-coupling phase of $q$-deformed chiral Yang-Mills theory is a
 topological string theory and its counting of branched covering maps
 is related to the counting of worldsheet instantons in the
 topological sigma-model. The phase structure of the gauge theory
 is similar to the familiar Crescimanno-Taylor picture, to which it is
 smoothly connected in the undeformed limit. The presence of a phase
 transition is also confirmed by the finite
 radius of convergence of the perturbative string expansion.

In the strong coupling phase the chiral deformed theory is related
 to an emerging toric geometry through the topological vertex formalism
of \cite{Aganagic:2003db}. The gauge theory thus captures the
 Calabi-Yau geometry with the appropriate D-brane insertions
 that are relevant for the counting of the black hole
microstates in the effective four dimensional supergravity theory. In this
 setup the ``distance'' between the fiber D-branes and the
 base sphere plays the role of a geometric K\"ahler modulus.
Remarkably, the large $N$ chiral $q$-deformed theory
 can be used efficiently to compute Gromov-Witten and Gopakumar-Vafa
 invariants of the toric geometry. On the other hand, the Gross-Taylor
 expansion of QCD$_2$ is recovered in a suitable double scaling
 limit. This expansion is known to compute Hurwitz numbers, the
 combinatorics of  branched coverings of the sphere. This connection
 is summarized by the explicit localization formula for the Gromov-Witten
invariants. Moreover, this relation
 can be exploited to study the analytic properties of the topological
string perturbation series. The existence of a finite radius of convergence
reflects the phase structure that we derived from a numerical analysis of the
 matrix model. Because the topological string is
  connected to the counting of BPS states, the divergence
 of its perturbation series could be physically related to
 a Hagedorn transition. A possibly
 related phase transition has been found recently in
 \cite{Papadodimas:2005wm}, where the topological $\mathcal{N}=4$
 gauge theory is studied which closely resembles the gauge theory
 defined on the $N$ D4-branes that localizes to the $q$-deformed
 Yang-Mills theory.

Together with the results in \cite{Caporaso:2005ta} we find a satisfactory
 and consistent confirmation of the ideas presented in
 \cite{Aganagic:2004js}. The deformed Yang-Mills theory that computes
 the BPS degeneracies of four-dimensional black holes is deeply
 related to the geometrical invariants
 of the relevant Calabi-Yau threefold. On the other hand, these
 are the objects underlying the topological string amplitudes that
 compute the four-dimensional effective field theory F-terms. The
 phase structure of the gauge theory is reflected in the topological
 string amplitude in a very precise fashion. The topological
 string theory also provides an explicit realization of the
 Gross-Taylor string expansion of the $q$-deformed Yang-Mills theory.

In principle, the techniques used in this paper can be extended to
compute Gromov-Witten invariants for the more general geometries of
\cite{Aganagic:2004js}. Similar issues were already
 addressed in \cite{Bryan:2004iq}. The chiral expansion of
$q$-deformed Yang-Mills theory could in general lead to a better
understanding of the relevant Calabi-Yau threefolds.

It would be fruitful to understand better the implications of the phase
transition for black holes physics. In the coupled expansion we encountered a
third-order phase transition of Gross-Witten type. Recently it has been
proposed \cite{Alvarez-Gaume:2005fv,Basu:2005pj}
that such behaviour could be related to a change of regime from a
macroscopic black hole to a perturbative string state, in the spirit of the
Horowitz-Polchinski transition \cite{Horowitz:1996nw}. On the other hand the
exponential growth in the density of BPS states observed here
should suggest some type of Hagedorn behaviour. It would be
interesting to find a string dual exhibiting the same growth of states at
the perturbative level, as in the correspondence between heterotic
strings on $\mathbb{T}^6$ and Type~IIA strings on
$K3\times\mathbb{T}^2$ \cite{Vafa:1995bm}, and to understand the phase
transition there as done in \cite{Dabholkar:2004yr}.

\acknowledgments

This work was supported in part by PPARC Grant PPA/G/S/2002/00478 and
by the EU-RTN Network Grant MRTN-CT-2004-005104.

\appendix
\addcontentsline{toc}{section}{Appendices}

\section{Degree zero partition function\label{Constmaps}}

In this appendix we will derive the contribution ${\sf Z}_0(q)$ of constant
maps to the topological string partition function by computing the
normalization of the $q$-deformed Yang-Mills partition function. For
this, we consider the quantity
\bea
S_{00}(q,N)&:=&\prod_{1\leq i<j\leq
  N}\,\left[q^{(j-i)/2}-q^{-(j-i)/2}\right]\nonumber\\
&=&\exp\biggl[\,-\sum_{1\leq i<j\leq N}\,\left(
\frac{j-i}{2}\,\log(q) - \log\bigl( 1
  - q^{j-i}\bigr)\right)  \biggr]   \ ,
\eea
where the second line holds up to an irrelevant phase factor.
The first term in the exponential can be easily summed according to
\beq
 \sum_{1\leq i<j\leq N}\,\frac{j-i}{2} =\frac{1}{2}\,
\sum_{i=1}^{N-1}\,\left( \frac{N \, (N+1)}{2} - \frac{i \, (i+1)}{2}
  +i^2 - N \, i \right) = \frac{N^2 \,\left(N-1\right)}{12} \ .
\eeq
In the second term, since $ q = \e^{-g_s}\in(0,1)$ and $j-i > 0  $ we
can expand the logarithm in powers of $q$ to obtain
\begin{eqnarray}
 \sum_{1\leq i<j\leq N}\,\log \big( 1 - q^{j-i}\big) &=& -
\sum_{m=1}^{\infty}~\sum_{i=1}^{N-1} \, \frac{q^{-i \, m}}{m}\,
\frac{ q^{m \, (N+1)} -  q^{m \, (i+1)}  }{q^m - 1} \\
 &=& - \sum_{m=1}^{\infty} \, \left(
\frac{N \, q^m}{m \, (1-q^m)} -\frac{q^m}{m \, (1-q^m)^2} + \frac{q^{m
    \, (N+1)}}{m \, (1-q^m)^2} \right) \ . \nonumber
\label{2ndtermexp}\end{eqnarray}
The first two terms in (\ref{2ndtermexp}) can be rewritten in more
familiar forms by expanding again in powers of $q$ to get
\beq
 - \sum_{m=1}^{\infty} \, \frac{N \, q^m}{m \, (1-q^m)}= - N\,
\sum_{m=1}^{\infty}~\sum_{n=1}^{\infty}\,\frac{q^{m \, n}}{m}
= N \, \log\bigl(\eta(q)\bigr) - \frac{N}{24} \, \log(q)
\eeq
where
\beq
\eta (q) = q^{1/24}~\prod_{n=1}^{\infty}\,\left(1-q^n\right)
\eeq
is the Dedekind function, and similarly
\beq
 \sum_{m=1}^{\infty} \, \frac{q^m}{m \, (1-q^m)^2}=
\sum_{m=1}^{\infty}~\sum_{n=1}^{\infty}\,\frac{n}{m}~q^{m \, n}=
\log\bigl(M(q)\bigr)
\eeq
where
\begin{equation}
M(q) = \prod_{n=1}^\infty\,\frac1{\left(1-q^n\right)^{n}}
\label{McMahon}\end{equation}
is the McMahon function. The series expansion of (\ref{McMahon}) is
the generating function $\sum_{\cal Y}\,q^{|{\cal Y}|}$ for plane
partitions $\cal Y$ which can be used to define the topological vertex
of Section~\ref{toric}.

Collecting everything together we arrive at
\bea
\log\bigl(S_{00}(q,N)\bigr)&=&- \mbox{$\frac{1}{24}$}\,\bigl(2N^2 \,
(N-1)+N\bigr)\, \log(q) + N \, \log\bigl(\eta(q)\bigr)
\nonumber\\ &&+\,\log\bigl(M(q)\bigr) + \log\bigl(N_0 (q,Q)\bigr) \ ,
\eea
where we have introduced the function
\begin{equation}
  N_0(q,Q) = \exp\biggl(\,- \sum_{n=1}^{\infty}\,B_n(q)~Q^n\biggr)
\label{N0qQ}\end{equation}
with $Q:=q^N=\e^{-g_sN}$ and
\beq
B_n(q) = \frac{q^n}{n \, (1-q^n)^2} \ .
\eeq
The function $N_0(q,Q)$ encodes the nonperturbative corrections to the
degree zero map contribution, which are immaterial in the large $N$
limit. We may therefore factor it out of the partition function, after
which we arrive at the result (\ref{Z0qdef}) reported in the main
text.

The corresponding free energy
\beq
{\sf F}_0(q)=\lim_{N\to\infty}\,\log\left(\mbox{$
\frac{S_{00}(q,N)}{N_0(q,Q)}$}\right)
\label{freedeg0}\eeq
admits a genus expansion ${\sf F}_0=\sum_{g\geq0}\,g_s^{2g-2}~{\sf
  F}^{~}_0{}^g$. Its coefficients reproduce the degree zero Gromov-Witten
invariants $\hat{\rm N}^g{}^{~}_{(0,0,0)}(p)={\rm N}^g{}^{~}_0(p)$
which can be expressed entirely in terms of classical intersection
indices and characteristic classes of the threefold $X_p$ as
\bea
{\rm N}^0{}^{~}_{0}(p)&=&\frac1{3!}~\sum_{\alpha_{a_1},\alpha_{a_2},
\alpha_{a_3}\in H^*(X_p,\zed)}~\int_{X_p}\alpha_{a_1}\wedge
\alpha_{a_2}\wedge\alpha_{a_3} \ , \nonumber\\[5pt]
{\rm N}^1{}^{~}_{0}(p)&=&-\frac1{24}\,\int_{X_p}\tau\wedge c_2(X_p) \
, \nonumber\\[5pt]
{\rm N}^g{}^{~}_{0}(p)&=&\frac{(-1)^g\,|B_{2g}\,B_{2g-2}|}
{4g\,(2g-2)\,(2g-2)!}\,\int_{X_p}\bigl(c_3(X_p)-c_1(X_p)\wedge
c_2(X_p)\bigr)
\label{GWdeg0}\eea
with $g\geq2$. Here $\{\alpha_a\}$ is a basis of $H^*(X_p,\zed)$
modulo torsion, the class $\tau$ is the degree two cohomology
generator and $c_n(X_p)$ is the $n$-th Chern class of the tangent bundle
$TX_p$. The coefficients $B_{2n}\in\rat$ are the Bernoulli numbers.

\section{Genus one partition function\label{Genus1partfn}}

In this appendix we collect the first five contributions to the genus
one partition function. Let us parameterize the genus one free energy as \beq
{\cal F}_1\big(t\,,\,\hat t\,;\,p\big)=
{{\left( 1- \e^{-{2\,\hat t}}\,\right) }^2}~\sum_{n=1}^\infty\,
\e^{-n\,t}~T_n\big(\hat t\,;\,p\big) \ . \eeq Then the first five
coefficients are given by
\begin{eqnarray}
T_1&=&-\frac{1}{12} \ , \nonumber\\[5pt]
T_2&=&-\frac{\e^{-4\,\hat t}}{48} \,\Bigl[2+
6\,p - p^2 - 4\,p^3 - p^4 + 2~\e^{2\,\hat t}\,\left( 2 -
5\,p^2 + p^4 \right)\Bigr.
\nonumber\\
&&\Bigl.+~\e^{4\,\hat t}\,\left(2 - 6\,p - p^2 + 4\,p^3 - p^4
      \right)\Bigr] \ , \nonumber\\[5pt]
T_3&=&-\frac{\e^{-8\,\hat t}}{72}
\,\Bigl[2 + 14\,p + 19\,p^2 - 20\,p^3 - 45\,p^4 - 24\,p^5 - 4\,p^6
\Bigr.\nonumber\\
     && \left.+\,
      6~\e^{4\,\hat t}\,\left(1 - 9\,p^2 + 13\,p^4 - 4\,p^6 \right)
\right.\nonumber\\
     && \left.+~
      \e^{8\,\hat t}\,\left(2 - 14\,p + 19\,p^2 + 20\,p^3 - 45\,p^4 +
 24\,p^5 - 4\,p^6 \right)\right.\nonumber\\
     && \left.+\,
      2~\e^{6\,\hat t}\,\left( 2 - 7\,p - 14\,p^2 + 34\,p^3 + 3\,p^4 -
        24\,p^5 + 8\,p^6 \right)\right.\nonumber\\
     && \Bigl.+\,
      2~\e^{2\,\hat t}\,\left( 2 + 7\,p - 14\,p^2 - 34\,p^3 + 3\,p^4 +
        24\,p^5 + 8\,p^6 \right)  \Bigr] \ , \nonumber\\[5pt]
T_4&=&-\frac{
\e^{-12\,\hat t}}{288}\,\Bigl[6 + 70\,p + 227\,p^2 + 80\,p^3 -
 744\,p^4 - 1332\,p^5 - 950\,p^6 - 312\,p^7 - 39\,p^8\Bigr.\nonumber\\
     && \left.+~
      \e^{12\,\hat t}\,\left(6 - 70\,p + 227\,p^2 - 80\,p^3 -
 744\,p^4 + 1332\,p^5 - 950\,p^6 + 312\,p^7 - 39\,p^8 \right)
\right.\nonumber\\
     && \left.+\,
      2~\e^{10\,\hat t}\,\left( 6 - 44\,p - 12\,p^2 + 452\,p^3 - 543\,p^4 - 438\,p^5 + 1080\,p^6 - 624\,p^7 + 117\,p^8 \right)\right.\nonumber\\
     && \left.+\,
      2~\e^{2\,\hat t}\,\left( 6 + 44\,p - 12\,p^2 - 452\,p^3 -
        543\,p^4 +438\,p^5 + 1080\,p^6 + 624\,p^7 + 117\,p^8 \right)
      \right.\nonumber\\
     && \left.+\,
      4~\e^{6\,\hat t}\,\left( 6 - 118\,p^2 + 471\,p^4 - 560\,p^6 +
195\,p^8 \right)\right.\nonumber\\
     && \left.+~
      \e^{8\,\hat t}\,\left(18 - 70\,p - 327\,p^2 + 832\,p^3 +
        888\,p^4 - 2244\,p^5 - 90\,p^6 + 1560\,p^7 - 585\,p^8 \right)
\right.\nonumber\\
     && \Bigl.+~
      \e^{4\,\hat t}\,\left(18 + 70\,p - 327\,p^2 - 832\,p^3 +
        888\,p^4+ 2244\,p^5 - 90\,p^6 - 1560\,p^7 - 585\,p^8 \right)
\Bigr] \ , \nonumber\\[5pt]
T_5&=&-\frac{\e^{-16\,\hat t}}{
    1440}\,\Bigl[24 + 404\,p + 2182\,p^2 + 3660\,p^3 + 5865\,p^4 -
 31604\,p^5 - 50870\,p^6\Bigr.\nonumber\\
    &&\left.-\,42200\,p^7 -
      19435\,p^8 - 4720\,p^9 - 472\,p^{10} +~\e^{10\,\hat t}\,
       \left(96 - 404\,p - 3242\,p^2 + 8620\,p^3\right.\right.
\nonumber\\ &&\left.\left. + 21470\,p^4 - 50176\,p^5 - 36880\,p^6 +
 101800\,p^7 -1540\,p^8 - 66080\,p^9 + 26432\,p^{10} \right)
\right.\nonumber\\ &&\left. +~
      \e^{16\,\hat t}\,\left(24 - 404\,p + 2182\,p^2 - 3660\,p^3 -
5865\,p^4 + 31604\,p^5 - 50870\,p^6 + 42200\,p^7
      \right.\right.\nonumber\\ &&\left.\left.-\,19435\,p^8 + 4720\,p^9-
         472\,p^{10} \right)  +2~\e^{14\,\hat t}\,\left( 24 - 278\,p +
         397\,p^2 + 3990\,p^3 - 13475\,p^4 \right.\right.
         \nonumber\\ &&\left.\left.+\, 5208\,p^5 + 32360\,p^6 - 57340\,p^7 +
         41410\,p^8 - 14160\,p^9 + 1888\,p^{10} \right)+2~\e^{2\,\hat t}\,
       \bigl( 24\bigr.\right.\nonumber\\ &&\left.\left.+\,
278\,p + 397\,p^2 - 3990\,p^3 - 13475\,p^4 - 5208\,p^5 + 32360\,p^6
+ 57340\,p^7 + 41410\,p^8
       \right.\right.\nonumber\\ &&\left.\left. +\, 14160\,p^9 +
       1888\,p^{10} \right)  +
      10~\e^{8\,\hat t}\,\left(12 - 418\,p^2 + 3257\,p^4 - 8810\,p^6+
 9275\,p^8\right.\right.\nonumber\\
      &&\left.\left.-\,3304\,p^{10} \right)  +
      4~\e^{12\,\hat t}\,\left(18 - 139\,p - 311\,p^2 + 3010\,p^3 -
 1235\,p^4 - 13436\,p^5 \right.\right.\nonumber\\
      &&\left.\left.+\,16770\,p^6 + 9180\,p^7 - 27055\,p^8 +
 16520\,p^9 -3304\,p^{10} \right)  + 4~\e^{4\,\hat t}\,\bigl(18 + 139\,p
\bigr.\right.\nonumber\\
         &&\left.\left.-\,311\,p^2 - 3010\,p^3 - 1235\,p^4 +
 13436\,p^5 + 16770\,p^6 - 9180\,p^7 -
         27055\,p^8\right.\right.\nonumber\\ &&\left.\left.-\,
16520\,p^9 - 3304\,p^{10} \right)  + 2~\e^{6\,\hat t}\,
       \left( 48 + 202\,p - 1621\,p^2 - 4310\,p^3 + 10735\,p^4
\right.\right.\nonumber\\
       &&\Bigl.\left. +\,25088\,p^5 - 18440\,p^6 - 50900\,p^7 -
         770\,p^8 + 33040\,p^9 + 13216\,p^{10} \right) \Bigr] \ .
\end{eqnarray}

For the analysis of Section~\ref{Genus1} we require only the leading
order expansion in $p$ given by
\begin{eqnarray}
{\cal F}_1^\infty\big(t\,,\,\hat t\,;\,p\big)&=&\mbox{$
 \frac{1}{48}$}\,{\left( 1 - \e^{-2\,\hat t}\,\right)}^4~
\e^{-2\,t}\,p^4 +\mbox{$
  \frac{1}{18 }$}\,{\left( 1 - \e^{-2\,\hat t}\,\right) }^6~
\e^{-3\,t}\,p^6\nonumber\\
  &&+\,\mbox{$\frac{13}{96}$}\,{\left( 1 - \e^{-2\,\hat t}\,\right) }^8~
\e^{-4\,t}\,p^8 +\mbox{$
  \frac{59}{180}$}\,{\left( 1 - \e^{-2\,\hat t}\,\right) }^{10}~
\e^{-5\,t}\,p^{10} +\cdots\nonumber\\
  &=& \sum_{n=1}^\infty\,\frac{R(n)}{(2 n)!}\,{{\left( 1 -
\e^{-2\,\hat t}\,\right) }^{2 n}~\e^{-n\,t}\,p^{2 n}} \ .
\end{eqnarray}
The coefficients $R(n)$ corresponding to the different winding numbers are
recognized to be \beq R(n)=\frac{(2 n)!}{24\, n!}\,\left(n^n - n^{n
- 1} -
      \sum_{k=2}^n\,(k - 2)!\,{n\choose k}\,n^{n - k}\right) \ .
\eeq They count the number of branched coverings of the sphere by a torus
with simple ramification corresponding to the trivial partition $(1^n)$.

The asymptotic behaviour of $R(n)$ for large $n$ can be determined by
means of the integral representation \bea
H(n):=\frac{R(n)}{(2n)!}&=&\frac{1}{24\, n!}\,\left(n^n - n^{n
- 1} -\int^\infty_0 \dd t~
\sum_{k=2}^n\,t^{k-2}~\e^{-t}\,{n\choose k}\,n^{n - k}\right)\nonumber\\
     &=&\frac{1}{24\, n!}\,\left(n^n - n^{n - 1} -n^n\,
     \int^\infty_0  \frac{\dd t}
    {t^2}~\e^{-t}\,\left[ -1 - t +
      {\left( \frac{n + t}{n} \right) }^n \right] \right)\nonumber\\
      &=&\frac{1}{24\, n!}\,\left(n^n - n^{n - 1} -
     \int^\infty_0  {\dd t}~\e^{-t}\,
{\left[n^n- {\left( {n + t} \right) }^{n-1} \right] }\right)\nonumber\\
 &=&\frac{1}{24\, n!}\,\left(~
     \int^\infty_0  {\dd t}~\e^{-t}\,
{ {\left( {n + t} \right) }^{n-1}  }-n^{n - 1}\right) \ . \eea This
last integral can be computed explicitly, giving a closed form
for these combinatorial numbers in terms of the incomplete
gamma-function as \beq
H(n)=\frac{1}{24\, n!}\,\Bigl(\e^n\,\Gamma(n, n)-n^{n - 1}\Bigr)=
\frac{1}{24\, n!}\,\left(\frac{\e^n\,\Gamma(n+1,n)}{n}-2\,n^{n -
1}\right)\ . \eeq By using the asymptotic large $n$ expansion \beq
\Gamma(n+1,n)=\e^{-n}\,
n^n\,\left(\,\mbox{$\sqrt{\frac{\pi}{2}}\,n^{1/2}+\frac{2}{3}+
\frac{\sqrt{2\pi}}{24}\,n^{-1/2}+\cdots$}\right)
\eeq along with the Stirling approximation we obtain \beq
\lim_{n\to\infty}\,H(n)=\frac{n^{n-1/2}\,\sqrt{\frac{\pi }{2}}}{24\, n!}
~\stackrel{n\to\infty}{\longrightarrow}~
\frac{ n^{n-1/2}\,\sqrt{\frac{\pi }{2}}}{24\,
{n^{n+1/2}\,{\sqrt{2\pi }}}~\e^{-n}} =\frac{\e^n}{48 n} \ .
\eeq

\section{Gopakumar-Vafa invariants\label{GopVafainvs}}

In this appendix we list the genus zero Gopakumar-Vafa invariants
${\rm n}^0{}^{~}_n(p)=\hat{\rm n}^0{}^{~}_{(0,n,0)}(p)$ of the
threefold $X_p$ for $n=1,\dots,7$, all of which are computed by the
method described in Section~\ref{GWinvs}:
\bea
{\rm n}^0{}^{~}_1(p)&=& ( -1)^{p} \ , \nonumber\\[5pt]
{\rm n}^0{}^{~}_2(p)&=&\mbox{$\frac{1}{8}$}\,\left(1-(-1)^p-4\,p+2\,
p^2\right) \ , \nonumber\\[5pt]
{\rm n}^0{}^{~}_3(p)&=&-(-1)^p\,\left(
\mbox{$\frac{1}{3}\,p-\frac{5}{6}\,p^2+\frac{2}{3}\,p^3-
\frac{1}{6}\,p^4$}\right) , \nonumber\\[5pt]
{\rm n}^0{}^{~}_4(p)&=&-\left(\mbox{$\frac{1}{6}\,p-
  \frac{13}{12}\,p^2+\frac{7 }{3}\,p^3 - \frac{9}{4}\,p^4 +p^5-
\frac{1}{6}\,p^6$}\right) \ , \nonumber\\[5pt]
{\rm n}^0{}^{~}_5(p)&=&-(-1)^p\,\left(\mbox{$\frac{1}{6}\,p -
  \frac{5}{4}\, p^2+
  \frac{9}{2}\,p^3 -
  \frac{69}{8}\,p^4+
  \frac{55}{6}\,p^5 -
  \frac{65}{12}\,p^6 +
  \frac{5}{3}\,p^7 -
  \frac{5}{24}\,p^8$}\right) \ , \nonumber\\ [5pt]
{\rm n}^0{}^{~}_6(p)&=&-\left(\mbox{$
\frac{13}{120}\,p -\frac{{\left( -1 \right) }^p}{24}\,p -
  \frac{313}{240}\,p^2 +\frac{5\,{\left( -1 \right) }^p}
   {48}\,p^2 + \frac{83}{12}\,p^3 -
  \frac{{\left( -1 \right) }^p}{12}\,p^3 - \frac{333}{16}\,p^4
$}\right. \nonumber\\ && \left. +\,\mbox{$
  \frac{{\left( -1 \right) }^p}{48}\,p^4 + \frac{757}{20}\,p^5 -
  \frac{1025}{24}\,p^6 + 30\,p^7 - \frac{51}{4}\,p^8 + 3\,p^9 -
  \frac{3}{10}\,p^{10}$}\right) \ , \nonumber\\[5pt]
{\rm n}^0{}^{~}_7(p)&=&-(-1)^p\,\left(\mbox{$
\frac{1}{10}\,p - \frac{241}{180}\,p^2 + \frac{851}{90}\,p^3 -
 \frac{7163}{180}\,p^4 + \frac{38269}{360}\,p^5 -
 \frac{134407}{720}\,p^6$}\right. \nonumber\\ && \left.
+\,\mbox{$\frac{19747}{90}\,p^7 -
  \frac{24941}{144}\,p^8 + \frac{6517}{72}\,p^9 -
 \frac{2401}{80}\,p^{10} + \frac{343}{60}\,p^{11}
 - \frac{343}{720}\,p^{12}$}\right) \ .
\eea
Note that for $p=0,1,2$ one has ${\rm n}^0{}^{~}_n(p)=0$ for $n\neq 1 $,
and by using eq.~(\ref{GWGV0rel}) one finds that the corresponding
Gromov-Witten invariants are given by ${\rm
  N}^0{}^{~}_{n}(p)=(-1)^p/n^3$. This is the expected result for
smoothly embedded contractible rational curves~\cite{bryankatzleung}.
For $p\geq3$ the structure changes. For example, at $p=3$ one finds ${\rm
  n}^0{}^{~}_7(3)\neq0$.

\section{Chiral integrals\label{ChiralInts}}

The saddle-point solution of the chiral $q$-deformed gauge theory
in the large $N$ limit requires the elementary indefinite integral \bea &&
\int \frac{\dd w}{w-s}~\frac{1}{\sqrt{(w-\e^{c^\prime})(w-\e^{b^\prime})}}\\
&&\qquad\qquad~=~-\frac{1}{\sqrt{(s-\e^{c^\prime})(s-\e^{b^\prime})}}\,
\log\left(\mbox{$\frac{\left(\sqrt{(w-\e^{b^\prime})(s-\e^{c^\prime})}+
\sqrt{(s-\e^{b^\prime})(w-\e^{c^\prime})}~\right)^2}{(s-w)\,
\sqrt{(s-\e^{b^\prime})(s-\e^{c^\prime})}}$}\right) \nonumber
\eea in the complex plane. The cuts in both $s$ and $w$ are taken as
indicated in Fig.~\ref{fg}.
Then one has \bea && \int_{-\infty}^{-\epsilon}\,\frac{\dd
w}{w(w-s)}~\frac{1}{\sqrt{(\e^{c^\prime}-w)(\e^{b^\prime}-w)}}\\
&& \qquad ~=~
\frac{1}{s}\,\left(~\int_{-\infty}^{-\epsilon}\,\frac{\dd w}{w-s}~
\frac{1}{\sqrt{(\e^{c^\prime}-w)(\e^{b^\prime}-w)}}
-\int_{-\infty}^{-\epsilon}\,\frac{\dd w}{w}~
\frac{1}{\sqrt{(e^{c^\prime}-w)(e^{b^\prime}-w)}}\right)\nonumber\\
&& \qquad ~=~
\frac{1}{s\,\sqrt{(s-\e^{b^\prime})(s-\e^{c^\prime})}}\,
\log\left(\mbox{$\frac{\left(\e^{{b^\prime}/{2}}\,\sqrt{s-\e^{c^\prime}}
+\e^{{c^\prime}/{2}}\,\sqrt{s-\e^{b^\prime}}~\right)^2}{s\,\left(\sqrt{s-\e^{c^\prime}}
+\sqrt{s-\e^{b^\prime}}~\right)^2}$}\right)\nonumber\\
&& \qquad\qquad\qquad+\,\frac{\e^{-(b'+c'\,)/2}}{s}\,
    \left[ b' + {c'} - 2\,\log\left(\mbox{$\frac{\e^{{b'}/{2}} +
\e^{{c'}/{2}}}2$}\,\right) - \log (\epsilon ) \right] \ . \nonumber
\eea The integral over the cut $[\e^{c^\prime},\e^{d^\prime}]$ is
instead given by \bea &&
\int_{\e^{c^\prime}}^{\e^{d^\prime}}\,\frac{\dd w}{w(w-s)}~
\frac{1}{\sqrt{(\e^{c^\prime}-w)(\e^{b^\prime}-w)}}\\
&& \qquad ~=~\frac{1}{s}\,
\left(~\int_{\e^{c^\prime}}^{\e^{d^\prime}}\,\frac{\dd w}{w-s}~
\frac{1}{\sqrt{(\e^{c^\prime}-w)(\e^{b^\prime}-w)}}
-\int_{\e^{c^\prime}}^{\e^{d^\prime}}\,\frac{\dd w}{w}~
\frac{1}{\sqrt{(\e^{c^\prime}-w)(\e^{b^\prime}-w)}}\right)\nonumber\\
&& \qquad ~=~-\frac{1}{s\,\sqrt{(s-\e^{c^\prime})(s-\e^{b^\prime})}}\,
\log\left(\mbox{$\frac{\left(\sqrt{(\e^{d^\prime}-\e^{b^\prime})(s-\e^{c^\prime})}+
\sqrt{(s-\e^{b^\prime})(\e^{d^\prime}-\e^{c^\prime})}~\right)^2}
{\left(s-\e^{d^\prime}\right)\left(\e^{c^\prime}-\e^{b^\prime}\right)}$}\right)
\nonumber\\
&& \qquad\qquad\qquad -\,
\frac{\e^{-{(b^\prime+c^\prime\,)}/{2}}}s\,
\log\left(\mbox{$\frac{\left(\e^{{c^\prime}/{2}}\,\sqrt{\e^{d^\prime}-\e^{b^\prime}}+
\e^{{b^\prime}/{2}}\,\sqrt{\e^{d^\prime}-\e^{c^\prime}}~\right)^2}
{\e^{d^\prime}\,\left(\e^{c^\prime}-\e^{b^\prime}\right)}$}\right) \
. \nonumber
\eea

\end{document}